\RequirePackage{lineno}
\documentclass[a4paper,12pt]{article}
\usepackage[utf8]{inputenc}
\usepackage{amsmath,amssymb,textcomp}
\usepackage{graphicx}
\usepackage{color}
\usepackage{epsfig,authblk}
\usepackage{bm,fullpage}
\usepackage{abstract}
\usepackage{setspace}

\newcommand\T{\rule{0pt}{5ex}}
\newcommand\B{\rule[-3ex]{0pt}{0pt}}

\modulolinenumbers[2]

\title{Evolutionary comparison between viral lysis rate and latent period}
\author{Juan A. Bonachela\footnote{jabo@princeton.edu}}
\author{Simon A. Levin}
\affil{Department of Ecology and Evolutionary Biology,   Princeton University, Princeton, New Jersey, 08544, USA}
\date{}

\begin{document}

\hyphenchar\font=-1

\sloppy

\linenumbers

\doublespacing

\maketitle

\vspace*{0cm}

\newpage 

\begin{abstract}
Marine viruses shape the structure of the microbial community. They are, thus, a key determinant of the most important biogeochemical cycles in the planet. Therefore, a correct description of the ecological and evolutionary behavior of these viruses is essential to make reliable predictions about their role in marine ecosystems. The infection cycle, for example, is indistinctly modeled in two very different ways. In one representation, the process is described including explicitly a fixed delay between infection and offspring release. In the other, the offspring are released at exponentially distributed times according to a fixed release rate. By considering obvious quantitative differences pointed out in the past, the latter description is widely used as a simplification of the former. However, it is still unclear how the dichotomy ``delay versus rate description'' affects long-term predictions of host-virus interaction models. Here, we study the ecological and evolutionary implications of using one or the other approaches, applied to marine microbes. To this end, we use mathematical and eco-evolutionary computational analysis. We show that the rate model exhibits improved competitive abilities from both ecological and evolutionary perspectives in steady environments. However, rate-based descriptions can fail to describe properly long-term microbe-virus interactions. Moreover, additional information about trade-offs between life-history traits is needed in order to choose the most reliable representation for  oceanic bacteriophage dynamics. This result affects deeply most of the marine ecosystem models that include viruses, especially when used to answer evolutionary questions.
\end{abstract}

\newpage
\section*{Introduction}
Viruses are the most numerous organisms on Earth.
They play diverse roles in the biotic component of practically any ecosystem. Especially remarkable is the case of marine ecosystems. Marine viruses are important sources of mortality at every trophic level. Potential hosts range from whales and commercial fish species to zooplankton, heterotrophic bacteria and microbial autotrophs \cite{review0}. Viruses are key components of the microbial loop and, therefore, the biogeochemical cycle of elements such as nitrogen or phosphorus \cite{review}. They are responsible for more than $40\%$ of marine bacterial mortality \cite{review}, contributing importantly to shaping the community \cite{cyanos1,KtW,review1}. The relevance of {\it virioplankton} not only stems from the ``predatory'' pressure they exert, but also from the subsequent release of organic nutrients (able to supply a considerable amount of the nutrient demand of, e.g. heterotrophic bacterioplankton \cite{recycling}); or their contribution to microbial genetic diversity in the ocean through horizontal gene transfer \cite{review1,JMartiny,review2}.

\vspace{0.25cm}
The vast majority of these roles are assumed by marine viruses that eventually kill the host cell \cite{lytic_maj}. The standard {\it lytic infection} can be summarized in the following steps \cite{review3}: {\it i)} free viruses diffusing in the medium encounter and attach to cells at a certain {\it adsorption rate}; {\it ii)} after injecting its nucleic acid into the host cell, the virus takes control of the host synthesis machinery in order to replicate its genetic material (DNA or RNA, depending on the type of virus \cite{review2}) and produce the proteins that will form the components of the viral offspring ({\it eclipse period}); {\it iii)} during the {\it maturation} stage (or {\it rise period}), the new {\it virions} are assembled; {\it iv)} finally, the virus synthesizes the {\it holin} protein, which perforates the plasma membrane allowing viral endolysins (lysoenzymes) to reach and lyse the cell wall, thereby releasing offspring and cellular organic compounds to the medium.

\vspace{0.25cm}
The {\it latent period} (steps {\it ii}-{\it iv} above), controlled by the so-called {\it gene t} (or holin gene) \cite{gene_t}, is one of the most important viral life-history traits. So are the {\it burst size} (offspring number, intimately related to the duration of the infection), and the adsorption rate. The latent period is studied intensively in the viral literature not only due to its ecological importance, but also owing to the small pleiotropic effect that its evolutionary change has on other phenotypic traits \cite{models_solution}.

\vspace{0.25cm}
\noindent
On the other hand, the latent period links ecological and evolutionary change, as mutations in this trait influence the demography of the population and the environment influences which latent periods are favored by selection \cite{gene_t,eco_evo_virus}, closing in this way an eco-evolutionary feedback loop \cite{eco_evo}. Furthermore, the short generation times and numerous offspring of viruses facilitate rapid evolution \cite{JMartiny2}, and a possible overlap between ecological and evolutionary timescales. All these factors provide evidence for the importance of using a proper description of the ecological interactions between virus and host in order to make reliable evolutionary predictions.

\vspace{0.25cm}
In the theoretical literature for marine viruses, mostly centered on viruses that infect bacteria ({\it bacteriophages}), host-virus interactions are represented in two different ways. One approach explicitly considers the latent period imposing a fixed delay between the adsorption and the release of the offspring \cite{levin_model}. In the other approach, new viruses are continuously released at a certain {\it lytic rate}, with cells that are simultaneously infected bursting at different post-infection times, exponentially distributed \cite{RM_stability,RM_middelboe}. Thus, in the delay model the survival of each and every infected cell is ensured up to an infection age that equals the fixed latent period, whereas survival responds to a probabilistic rule in the rate model. The latter can actually be seen as a simplification of the former that facilitates mathematical and computational analysis of the interactions. Indeed, the ecological outcome of the two approaches seems to be, a priori, qualitatively similar in spite of the  obvious difference in the timing of the infection \cite{RM_DM_comparison}. While in the delay model progeny show periods of no release (e.g. initial stages of viral culture experiments), in the rate model viral offspring are liberated at all times. However, little attention has been paid to quantifying thoroughly how these differences affect the long-term predictions by the two kinds of models. Here, we aim to fill this gap.

\vspace{0.25cm}
In this paper, we focus on the eco-evolutionary differences between the two approaches to the description of the lytic infection cycle. This comparison may prove very useful to assess the evolutionary consequences of the simplifying assumptions in these models, and therefore the long-term reliability of a whole group of different models for host-virus dynamics available in the literature. The rate-based approach is used to model not only diverse aspects of host-lytic virus interactions \cite{joshua_PNAS}, but also other types of viral infection cycles such as lysogeny \cite{lysogeny_ex} or shedding \cite{HIV}. In the latter, viruses continuously produce and release virions during the entire infection period. Some examples include filamentous phages, and viruses of an enormous importance for humans such as Ebola, SARS, smallpox, varicella-zoster virus, and HIV \cite{nowak}. In some retroviruses such as HIV, both burst and continuous production modes have actually been suggested \cite{HIV}. Thus, this question transcends purely technical matters such as model selection. Indeed, this study can potentially serve to compare the evolutionary strategies of a wide selection of viruses with very different infection cycles.

\vspace{0.25cm}
\noindent
As a model case, we use bacteriophages, due to their importance for biogeochemical cycles; it also allows us to resort to the extensive modeling bibliography available, in which the two approaches to the infection cycle are used. On the other hand, we consider mutations only in the holin gene, in order to isolate the effects of evolution on the key differentiating trait for the two strategies: the latent period (or, equivalently, lysis rate). Thus, we first present the two models for lytic infection. After briefly comparing them from an ecological perspective, we turn our attention to their evolutionary divergences. Under this framework, we discuss the ecological and evolutionary contrast between the two forms for the life-history {\it trade-off} between latent period and burst size that have been proposed in the literature. Finally, we comment on the implications of all the above for the descriptions of host-virus interactions in general, and marine bacteriophages in particular. This study will contribute to the reliability of long-term predictions regarding the interaction between a wide variety of viruses and their hosts.

\section{Modeling host--virus interactions}
\subsection{{\it Environment}}
In order to compare the two approaches to the infection cycle, we first set common idealized environmental conditions by using two-stage chemostats \cite{2stage_chem}.

\vspace{0.25cm}
Two-stage chemostats are basically composed of a continuous culture for bacterial hosts, coupled to a continuous culture of co-existing bacteria and viruses. A flow of nutrients from a fresh medium to the first chemostat facilitates bacterial growth, and a flow of ``fresh'' hosts from the first chemostat to the second chemostat allows for the development of the viral population. Finally, both virus and bacterial cells are washed out from the second chemostat at a certain rate. The described flows, which can loosely resemble e.g. the continuous passage or migratory events occurring in the mammalian gastrointestinal tract \cite{gut_ex}, enable a steady state for the overall system. From the perspective of marine bacteriophages, {\it quasi}-stationary conditions may be found in stratified waters where cyanobacteria, among the most common targets for virioplankton, dominate.

\vspace{0.25cm}
\noindent
Such a steady state is very convenient from the mathematical standpoint, as is the continuous source of hosts, which helps alleviate the oscillations that are frequently observed in standard predator-prey models \cite{2stage_chem} (see below). In addition, the continuous flow of uninfected hosts constitutes a relief for the bacterial population from the evolutionary pressure of the virus and, therefore, prevents bacteria from embarking on an otherwise expected co-evolutionary arms race \cite{eco_evo_virus,joshua_PNAS}. This allows us to focus on viral evolution only. Thus, two-stage chemostats provide a controlled environment whose conditions are easily reproducible in the laboratory; they also offer general results that can be adapted to other environments, as discussed below.

\vspace{0.25cm}
Lastly, the environmental parameters are chosen to avoid multiple infections (see table \ref{table} in Appendix \ref{AppA}), preventing in this way any kind of intra-cellular competition among viruses. 

\subsection{{\it Ecological analysis of the delay model (DM)}}
Let us study first the approach to lysis in which the individuals of the viral population release their offspring after exactly the same latent period. The model describing this cycle implements explicitly the delay between adsorption and burst. If $[C]$ represents the concentration of uninfected bacterial cells, $[V]$ the concentration of free viruses, and $[I]$ the concentration of infected bacteria, the dynamics of the interactions between host and virus can be modeled using the equations \cite{levin_model}:

\begin{eqnarray}
\dfrac{d[C](t)}{dt}&=&\mu\;[C]-k[C][V]-w[C]+w[C_{0}]\label{delayed_C_eq}\\
\dfrac{d[I](t)}{dt}&=&k[C][V]-\left(k[C]_{t-L}[V]_{t-L}\right)e^{-wL}-w[I]\label{delayed_I_eq}\\
\dfrac{d[V](t)}{dt}&=&\left(b\;k[C]_{t-L}[V]_{t-L}\right)e^{-wL}-k[C][V]-m[V]-w[V],\label{delayed_V_eq}
\end{eqnarray}

\noindent
where $\mu$ represents the growth rate of uninfected hosts; $k$, the adsorption rate; $m$, the viral mortality or decay rate; $w$, the washout rate; $L$ and $b$, the viral latent period and burst size, respectively; and the subscript $t-L$ indicates that the term is evaluated a lytic cycle (latent period) in the past.

\vspace{0.25cm}
\noindent
We initially consider a monomorphic viral population (i.e. all individuals share the same phenotype --or trait values). Thus, the first equation describes the dynamics of the host population as a result of growth (first term), infection events (second term), washout process (third term) and inflow of uninfected cells (fourth term). The second equation considers the dynamics of infected cells, whose number grows due to adsorption events (first term), and decreases due to dilution (last term), and lysis of cells (second term); the latter term is the result of correcting the number of cells that were infected $L$ time steps before, $k[C]_{t-L}[V]_{t-L}$, using the probability for those cells to survive dilution during that time ($e^{-wL}$ term) \cite{levin_model}. Likewise, the free virus population grows owing to those lysed cells (first term, number of lysis events times the burst size), and is reduced by adsorption (second term), natural mortality (third term) or dilution (last term).

\vspace{0.25cm}
\noindent
We assume a simple Monod formulation for the growth rate of bacteria, given by:

\begin{equation}
\mu([N])=\dfrac{\mu_{max}[N]}{[N]+K_{N}},
\label{Monod_equation} 
\end{equation}

\noindent
in which $mu_{max}$ is the maximum growth rate for the cell and $K_{N}$ its half-saturation constant (defined as the concentration at which the growth rate of the cell equals half its maximum). See table \ref{table} for units. Note that we use here the standard assumption that infected cells effectively allocate all their resources to viral production (i.e. $\mu_{I}\sim0$).

\vspace{0.25cm}
\noindent
To these equations, we must add the dynamics of the nutrient:

\begin{equation}
\dfrac{d[N]}{dt}=w\left([N_{0}]-[N]\right)-\mu([N])[C]/Y,
\label{nutrient_equation} 
\end{equation}

\noindent
where $[N_{0}]$ is the inflow of nutrient feeding uninfected host cells and $Y$ is a yield or efficiency parameter accounting for the efficiency for bacterial cells to transform uptake into growth.

\vspace{0.25cm}
The equations above can be solved for the stationary state, expected under chemostat conditions. Writing, for simplicity, $\overline{\mu}=\mu([N]_{st})$, we obtain a trivial solution for the virus-free configuration $[V]_{st}=[I]_{st}=0$, $[C]_{st}=w[C_{0}]/(w-\overline{\mu})$. This solution is always feasible (see Appendix \ref{AppB}). On the other hand, the non-trivial steady state is given by the expressions:

\begin{eqnarray}
\left[C\right]_{st}&=&\dfrac{w+m}{k\left(be^{-wL}-1\right)}\label{stat_delayed_C}\\
\left[I\right]_{st}&=&\dfrac{(\overline{\mu}-w)(w+m)}{k\;w}\dfrac{\left(1-e^{-wL}\right)}{\left(be^{-wL}-1\right)}+\left(1-e^{-wL}\right)[C_{0}],\label{stat_delayed_I}\\
\left[V\right]_{st}&=&\dfrac{\overline{\mu}-w}{k}+\dfrac{w[C_{0}]\left(be^{-wL}-1\right)}{w+m}\label{stat_delayed_V},
\end{eqnarray}

\noindent
feasible as long as $L<\ln(b)/w$ and $\overline{\mu}>w\left(1-[C_{0}]/[C]_{st}\right)$ (see Appendix \ref{AppB}). The stability of the trivial solution is determined by the basic reproductive number, $\mathcal{R}_{0}$. This observable is the expected number of secondary infections arising from a single individual in an equilibrium susceptible population \cite{newR0}. Thus, $\mathcal{R}_{0}<1$ indicates an eventual fall into the phage-free state , whereas $\mathcal{R}_{0}>1$ points to the instability of this trivial state. From Eq.(\ref{delayed_V_eq}), it is easy to find that $\mathcal{R}_{0}=\frac{b\;k\left[C\right]_{st}e^{-wL}}{k[C]_{st}+m+w}$. Thus, the trivial equilibrium changes stability for $\mathcal{R}_{0}=1$ or, in other words, for a $[C]_{st}$ given by Eq.(\ref{stat_delayed_C}). Therefore, the latter condition opens the possibility of a {\it transcritical bifurcation} \cite{AD_book}. However, the deduction of general stability conditions for the non-trivial state is a highly nontrivial task, beyond the scope of this article. We refer the reader to the extensive mathematical literature devoted to study of the local and global stability of host-virus systems similar to the one presented here \cite{DM_stability}. On the other hand, oscillations are a common outcome of predator-prey interactions, and frequently seen in bacteriophage models \cite{dushoff}. For the sake of mathematical tractability (especially for evolutionary matters), we focus our analysis on the region of the parameter space where stationarity can be found (but see Discussion).

\vspace{0.25cm}
\noindent
With these words of caution, we assume hereafter that the generic feasible steady state above fulfills those stability conditions and proceed with the rest of the analysis. Indeed, Eqs.(\ref{stat_delayed_C})-(\ref{stat_delayed_V}) prove to be stable for the realistic range of parameters used in this study, as shown in our simulations below. The chosen parametrization represents generically marine lytic T bacteriophages and a bacterial species (see table \ref{table}). 

\vspace{0.25cm}
Finally, the stationary growth rate for the viral population (per-capita change in the concentration of free virus) is given by:

\begin{equation}
\mu_{v} = \left(b\;e^{-wL}-1\right)k\;[C]_{st}.
\label{DM_fitness}
\end{equation}

\noindent
Because the average number of surviving offspring per cell is given by:

\begin{equation}
<b>=be^{-wL},
\label{av_b_delay}
\end{equation}

\noindent
Eq.(\ref{DM_fitness}) indicates that stationary co-existence is possible (i.e. $\mu_{v}=m+w$) only when $<b>=\frac{m+w}{k[C]}+1$ (i.e. $\mathcal{R}_{0}=1$), which ensures that the average number of offspring per cell is larger than one.

\subsection{{\it Ecological analysis of the rate model (RM)}}
Following the same notation above, infections in which the offspring are released at a certain lysis rate $k_{L}=1/L$ can be described by the equations:

\begin{eqnarray}
\dfrac{d[C](t)}{dt}&=&\mu\;[C]-k[C][V]-w[C]+w[C_{0}]\label{rate_C_eq}\\
\dfrac{d[I](t)}{dt}&=&k[C][V]-k_{L}[I]-w[I]\label{rate_I_eq}\\
\dfrac{d[V](t)}{dt}&=&b\;k_{L}[I]-k[C][V]-m[V]-w[V],\label{rate_V_eq}
\end{eqnarray}

\noindent
where the delay terms have been replaced by instantaneous terms (i.e. evaluated at time $t$). In this way, there is unceasingly new virions joining the free virus population, with host cells being lysed at a rate $k_{L}$. Eqs.(\ref{Monod_equation})-(\ref{nutrient_equation}) complete the description of the dynamics of the system.

\vspace{0.25cm}
The stationary states are described by the trivial configuration $[V]_{st}=[I]_{st}=0$, $[C]_{st}=w[C_{0}]/(w-\overline{\mu})$, always feasible (see Appendix \ref{AppB}), and the non-trivial steady state:

\begin{eqnarray}
\left[C\right]_{st}&=&\dfrac{\left(w+m\right)\left(k_{L}+w\right)}{k\left[k_{L}\left(b-1\right)-w\right]}\label{stat_rate_C}\\
\left[I\right]_{st}&=&\dfrac{(\overline{\mu}-w)(w+m)}{k\left[k_{L}\left(b-1\right)-w\right]}+\dfrac{w[C_{0}]}{k_{L}+w}\label{stat_rate_I}\\
\left[V\right]_{st}&=&\dfrac{\overline{\mu}-w}{k}+w[C_{0}]\dfrac{k_{L}\left(b-1\right)-w}{\left(w+m\right)\left(k_{L}+w\right)}\label{stat_rate_V},
\end{eqnarray}

\noindent
feasible for $L<(b-1)/w$ and $\overline{\mu}>w\left(1-[C_{0}]/[C]_{st}\right)$. The definition of the basic reproductive number, $\mathcal{R}_{0}=\frac{b\;k_{L}\left[I\right]_{st}}{\left(k[C]+m+w\right)\left[V\right]_{st}}$, leads once more to a potential transcritical bifurcation for $[C]_{st}$ given by Eq.(\ref{stat_rate_C}), which results from the condition $\mathcal{R}_{0}=1$. Also similarly to the previous model, our realistic parametrization is able to yield stable solutions. Thus, we assume that the stationary state fulfills the stability conditions and refer the reader to the existing literature for a detailed study of these \cite{RM_stability}. Lastly, we can deduce the viral growth rate as in the case of the DM:

\begin{equation}
\mu_{v} = \left(\dfrac{k_{L}b\;}{k_{L}+w}-1\right)k\;[C]_{st},
\label{RM_fitness}
\end{equation}

\noindent
for which, by realizing that:

\begin{equation}
<b>=\dfrac{k_{L}b}{k_{L}+w},
\label{av_b_rate} 
\end{equation}

\noindent
we can conclude again that the nontrivial stationary state is maintained thanks to the condition $<b>=\frac{m+w}{k[C]}+1$, again equivalent to $\mathcal{R}_{0}=1$.

\subsection{{\it The trade-off latent period -- burst size}}
It is advisable to note that latent period and burst size are not independent. The number of offspring is determined by the timing of the lysis. Moreover, the time spent in producing new virions increases the generation time (sum of extra- and intra-cellular viral lifetime). This sets an obvious life-history trade-off realized in the dichotomy ``immediate but low reproduction'' and ``delayed but larger offspring'' that shapes the evolution of $b$ and $L$.

\vspace{0.25cm}
Little information about life-history trade-offs is available for marine viruses. However, two mathematical forms have been suggested for this specific trade-off in the general bacteriophage literature. One takes into account that the parental virus is utilizing limited host resources to synthesize the virions \cite{exponential_tradeoff}:

\begin{equation}
b=\dfrac{M}{\gamma}\;\left(1-e^{-\gamma\;\left(L-E\right)}\right),
\label{exp_tradeoff}
\end{equation}

\noindent
where $M$ is the maturation rate, $E$ represents the eclipse period, and $\gamma$ is the decay rate for the bacterial resources. The other form assumes that the time needed to deplete host resources is much larger than the latent period, therefore simplifying the exponential relationship above to a linear function \cite{linear_tradeoff}:

\begin{equation}
b=M \left(L-E\right).
\label{lin_tradeoff}
\end{equation}

\noindent
Although the first option seems more mechanistic, most experimental evidence points to the linear relationship as the more frequently observed form for the trade-off \cite{linear_tradeoff}. In spite of the lack of information, and for the sake of concreteness, we assume here that marine phages can potentially show any of these two trade-offs. Thus, we perform all the calculations keeping in mind that $b$ is an increasing function of $L$, $f(L)$, replacing later such a function by each of the two forms mentioned above.

\section{Evolutionary Analysis}
Aiming to gain some knowledge on the evolutionary consequences linked to one or the other lytic descriptions, we now focus our attention on invasion experiments. Invasion analysis provides a unified framework with which we can reach some classic results, together with novel ones (see Appendix \ref{AppB}). As explained above, we consider only alterations on the gene {\it t}, controlling the duration of the latent period. Thus, mutants and residents differ only in $L$ (and, therefore, in $b$ as well). We consider that the form of the trade-off, $f(L)$, is the same for both viral populations.

\subsection{{\it An evolutionarily stable strategy (ESS) for the delay model}}
If we assume that the invading mutant (subscript $M$) perturbs the otherwise stable state of the resident population (subscript $R$), the possibility for invasion is decided by the sign of the invasion-matrix eigenvalue (see Appendix \ref{AppB}):

\begin{equation}
\lambda=\dfrac{1}{L_{M}}W_{n}\left(k[C_{R}]_{st}b_{M}L_{M}e^{-(k[C_{R}]_{st}+m)L_{M}}\right)-\left(k[C_{R}]_{st}+w+m\right),
\label{DM_eigen_sol_text} 
\end{equation}

\noindent
where $W_{n}(z)$ is the so-called Lambert function, defined as the solution to $W_{n}(z)e^{W_{n}(z)}=z$ \cite{lambert}. The analysis of the sign of $\lambda$ provides the condition for strategy $L^{*}$ to resist any invasion:

\begin{eqnarray}
f'\left(L^{*}\right)=w\;f\left(L^{*}\right).
\label{DM_der_sol_text}
\end{eqnarray}

\noindent
As shown in Appendix \ref{AppB}, the solution to this equation minimizes $[C]_{st}$ (Eq.(\ref{stat_delayed_C})) and maximizes $\mu_{v}$ (Eq.(\ref{DM_fitness})). Thus, $L^{*}$ is an ESS. This result is also reached after defining the invasion fitness function, $s_{L_{R}}(L_{M})=\lambda$, and analyzing its derivatives.

\vspace{0.25cm}
\noindent
The result of solving Eq.(\ref{DM_der_sol_text}) for the exponential trade-off, Eq.(\ref{exp_tradeoff}), and linear trade-off, Eq.(\ref{lin_tradeoff}), is summarized in table \ref{table0}. These results can be graphically obtained representing the pairwise invasibility plots (PIP) \cite{PIPs} for the two forms of $f(L)$. Fig.\ref{PIPs} (left) portrays the case of the exponential trade-off. 
 
\vspace{0.25cm}
\noindent
The sign of the invasion fitness depicted in the PIP provides essential information. Because $s_{L_{R}}(L_{M})<0$ for any $L_{M}$ when $L_{R}=L^{*}$, these solutions for the DM (table \ref{table0}) resist any invasion. $L^{*}$ also maximizes $\mu_{v}$, and therefore is an ESS. On the other hand, $s_{L_{R}}(L_{M})>0$ for $L_{R}<L_{M}<L^{*}$ and $L^{*}<L_{M}<L_{R}$, ensuring that phenotypes closer to $L^{*}$ can invade populations with phenotypes further from that strategy. Therefore, $L^{*}$ is also a convergence-stable strategy (CSS) \cite{PIPs}.

\begin{table}
\vskip4mm
\centering
\begin{tabular}{|c|c|c|}
\hline
Model&Trade-off&Evolutionarily Stable Strategy\\
\hline
\hline
DM&\begin{tabular}{c}Exponential\T\B\\\hline Linear\T\B\end{tabular}&\begin{tabular}{p{7cm}|p{5cm}}$L^{*}=\dfrac{1}{\gamma}\ln\left(1+\gamma/w\right)+E$&$b^{*}=M/(w+\gamma)$\T\B\\\cline{1-2}$L^{*}=1/w+E$&$b^{*}=M/w$\T\B\end{tabular}\\
\hline
\hline
RM&\begin{tabular}{c}Exponential\T\B\\\hline Linear\T\B\end{tabular}&\begin{tabular}{p{7cm}|p{5cm}}$L^{*}=\dfrac{-1}{\gamma}W_{n}\left(-e^{-\frac{\gamma\left(E w+1+w/\gamma\right)}{w}}\right)-\dfrac{\gamma+w}{\gamma w}$&$b^{*}=\dfrac{M}{\gamma}\;\left(1-e^{-\gamma\;\left(L^{*}-E\right)}\right)$\T\B\\\cline{1-2}\centering$-$&\centering$-$\T\B\end{tabular}\\
\hline
\end{tabular}
\caption{\footnotesize{Summary of ESS calculated analytically for the delay model (DM) and the rate-based model (RM) using the exponential and the linear forms for the trade-off between burst size $b$ and latent period $L$.}}
\label{table0}
\end{table}

\subsection{{\it An ESS for the rate model}}
After following similar steps to those of the previous section, the condition to be fulfilled for $L^{*}$ to be an ESS candidate is (see Appendix \ref{AppB}):

\begin{eqnarray}
f'\left(L^{*}\right)=\dfrac{w\;f\left(L^{*}\right)}{1+w\;L^{*}}.
\label{RM_der_sol_text}
\end{eqnarray}

\vspace{0.25cm}
\noindent
We can now combine Eq.(\ref{RM_der_sol_text}) with Eqs.(\ref{exp_tradeoff}) and (\ref{lin_tradeoff}) to obtain $L^{*}$ for the exponential and linear trade-offs, respectively. In the case of the former, the evolutionarily stable strategy can be found in table \ref{table0}.

\begin{figure}[h!]
\begin{center}
\includegraphics[width=7.5cm]{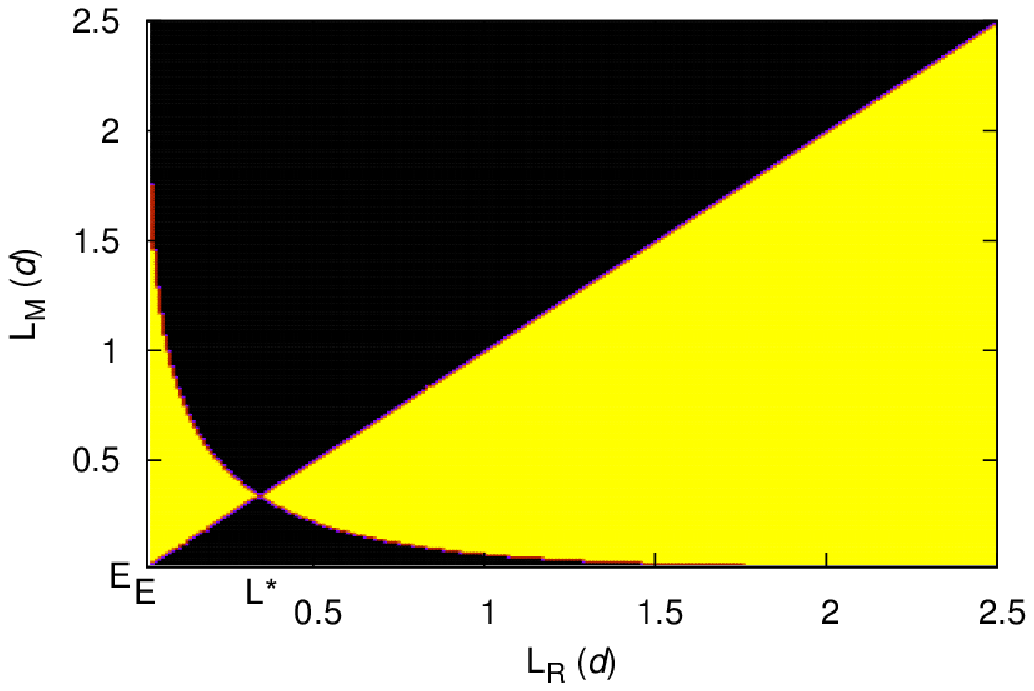}
\includegraphics[width=7.5cm]{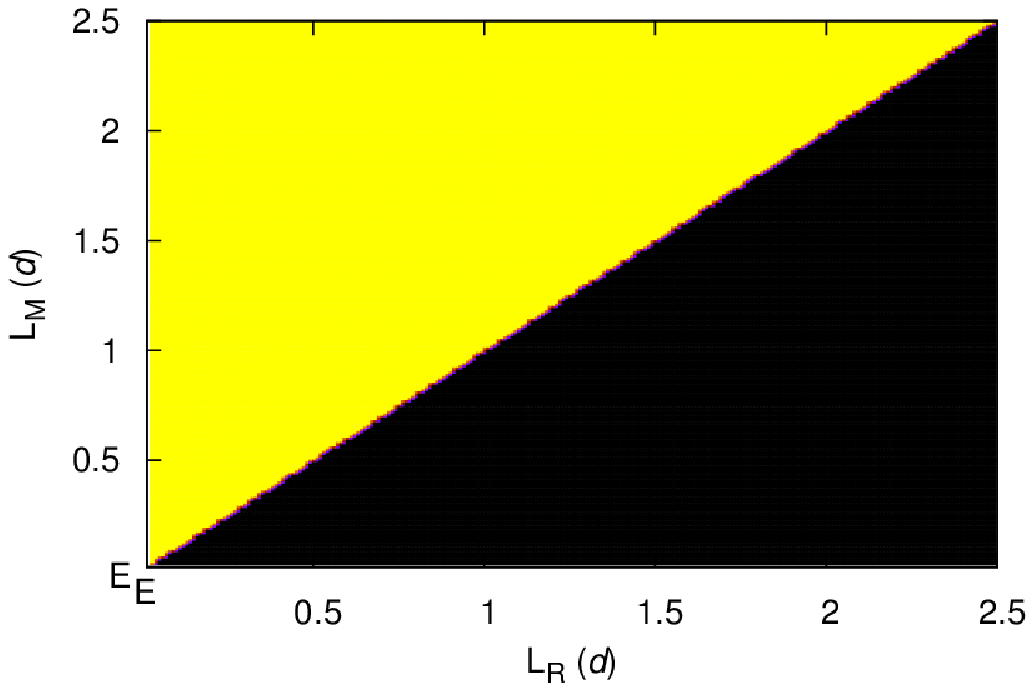}
\caption{\footnotesize{Pairwise invasibility plots. Yellow zones represent pairs of resident and mutant latent periods for which the mutant can take over the population (positive invasion fitness, $s_{L_{R}}(L_{M})$); in black zones, the resident can resist invasion (negative invasion fitness). Left: PIP obtained with the DM using the exponential form of the trade-off $b=f(L)$; no point on the vertical $L_{R}=L^{*}$ line falls into a yellow zone, ensuring that $L^{*}$ is an ESS. Qualitatively-similar results are found with the linear trade-off, and for the exponential $b=f(L)$ version of the RM. Right: PIP obtained with the RM using the linear form of the trade-off; the invasion fitness function only changes sign on the $L_{M}=L_{R}$ line and, therefore, no singular strategy is possible.}}
\label{PIPs}
\end{center} 
\end{figure}

This singularity maximizes fitness (see above) and, therefore, is an ESS. Furthermore, $L^{*}$ is, at least for the chosen parametrization, a CSS. These results can be easily confirmed, as in the previous case, with the numerical analysis of the derivatives of $s_{L_{R}}(L_{M})$ or plotting the corresponding PIP, qualitatively similar to the one shown in Fig.\ref{PIPs} (left).

\vspace{0.25cm}
\noindent
Interestingly, combining Eq.(\ref{lin_tradeoff}) with Eq.(\ref{RM_der_sol_text}) (or $\partial s/\partial L =0$) does not offer any feasible solution. Moreover, the only line providing a change of sign for the invasion fitness is the diagonal $L_{M}=L_{R}$ (see Fig.\ref{PIPs}, right panel).

\section{{\it Unconstrained evolution}}
Invasion analysis focuses on natural selection. The little pleiotropy expected for mutations affecting $L$, and the steady environment reached under chemostat conditions, allow invasion analysis to provide reliable evolutionary predictions if there are infinitesimal differences between mutant and parent phenotypes \cite{fitness_gradient}. These limitations are shared by most of the available theoretical frameworks, which may also require ecological equilibrium as necessary condition for mutation/immigration events to occur \cite{AD_book}. 

\vspace{0.25cm}
\noindent
We present now numerical simulations aimed to check if the ESSs calculated above are indeed reached in the absence of these constraints. To this end, we use an eco-evolutionary framework in which new mutants are created in the system at random times not necessarily coinciding with ecological stationary states. Similarly to the framework employed in \cite{joshua_PNAS,joshua_other}, new phenotypes can be introduced periodically in the system (invasion by migration), or by creating mutants through a genetic algorithm used at random times. These times are in part determined by each phenotype's population size and a common, fixed mutation rate. New phenotypes are identical to the mutating phenotype except for the latent period, which changes in an {\it arbitrarily} large amount.

\vspace{0.25cm}
The simulation scheme is basically as follows: the model equations (ecological interactions describing either the DM or the RM) are numerically integrated; at times calculated as specified above, a new (invading/mutant) virus phenotype is introduced into the system. The ecological dynamics then resumes, now with a new viral population in the medium competing against the existing ones (namely the dominant resident and contenders) for the host, which is their only available resource. This competition drives some phenotypes to extinction, which may change which species dominates, and thus the $(L,b)$ of the total viral population. These steps are repeated until, eventually, one phenotype comes along that is able to resist any invasion. This phenotype will remain as the dominant strain regardless of the other existing or incoming phenotypes. If this phenotype is indeed an ESS, its fitness will be larger than that of any other strain. Thus, due to its competitive advantage, the population size of the dominant phenotype will eventually be much larger than that of any other species in the system, and the average $(L,b)$ of the population will converge in the long term to this species' trait pair, $(L^{*},b^{*})$.

\begin{figure}[h!]
\begin{center}
\includegraphics[width=7.5cm]{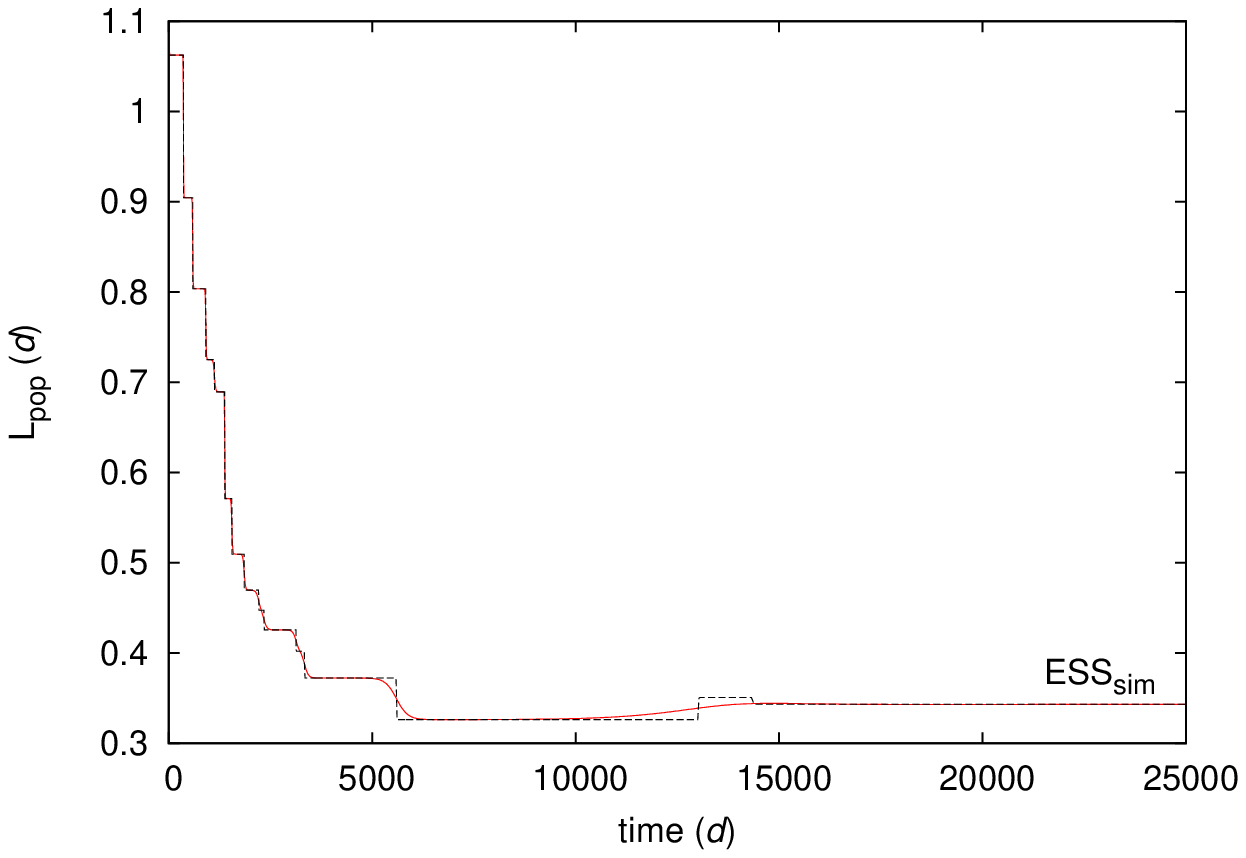}
\includegraphics[width=7.5cm]{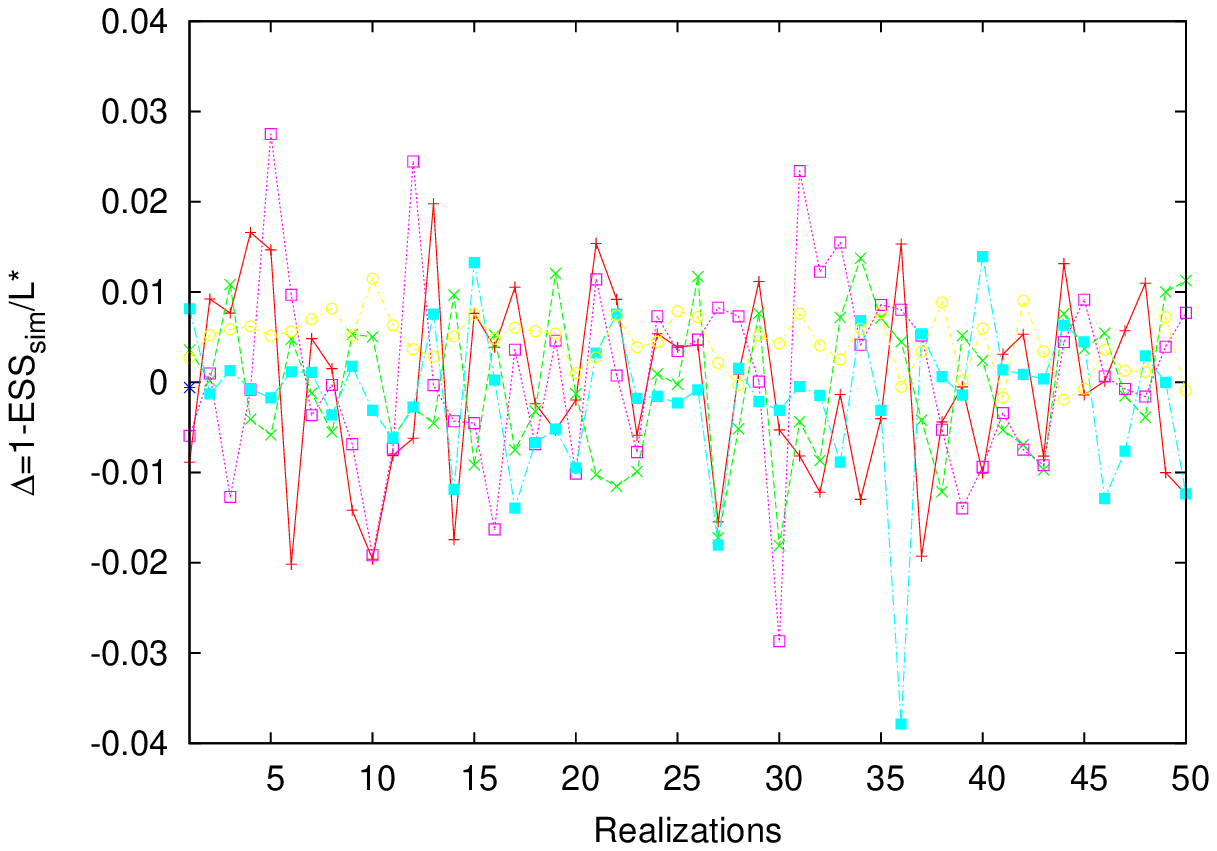}
\caption{\footnotesize{Eco-evolutionary simulation of bacteria-virus interactions for both descriptions. Left: Typical example of evolutionary succession in the DM (exponential trade-off); the dominant phenotype (black) changes with time due to mutation and selection, and with it the population latent period (red). Eventually, a strategy resisting any invasion (ESS$_{\textnormal{sim}}$) is reached. Right: Relative difference between the evolutionarily stationary value of the population latent period in simulations, ESS$_{\textnormal{sim}}$, and the analytic solution for each model, $L^{*}$. The evolutionary simulations shown here are those with the DM for exponential (red) and linear (green) trade-offs, the RM with exponential trade-off (blue), and their respective EiE counterparts (pink, cyan and yellow, respectively). The difference with the analytical result is never beyond $4\%$.}}
\label{sims}
\end{center} 
\end{figure}

\vspace{0.25cm}
The evolutionary succession described above can be observed in Fig.\ref{sims} (left). This figure portrays the average latent period of the population, as well as the latent period of the instantaneous dominant phenotype, over time. We keep in these simulations the parametrization given by table \ref{table}. After a long transient, the population reaches an evolutionarily stable state, ESS$_{\textnormal{sim}}$. As depicted in Fig.\ref{sims} (right), the evolutionary steady states obtained in different realizations of the framework fluctuate around the analytical ESS, with a very small variance. This result is robust, for it is observed either using random mutation times or periodic immigration events. Moreover, this result is also reached when an ``everything-is-everywhere'' (EiE) approach is used \cite{bruggeman,Follows}. In EiE approaches, a large number of fixed phenotypes, that is, with no possibility for evolutionary change, is used to initialize the system; these phenotypes, intended to represent all possible genetic variability, compete for the available resource until only one strain remains. As we observe in Fig.\ref{sims} (right), the phenotype able to out-compete the rest is consistently close to the ESS predicted analytically for the lytic strategy and trade-off used in the simulation. These results are observed for any combination of model/trade-off studied above except for the linear case of the RM for which, as deduced earlier, no ESS is expected.

\vspace{0.25cm}
In any case studied, an initial test simulation with monomorphic viral and host populations showed stationary values for $[C]$, $[V]$ and $[I]$ that match the analytical solutions, confirming the assumed stability of the stationary state.

\section{Discussion and comparison of models}

\subsection{{\it Ecological comparison}}
The steady-state value of the observables deduced above, and how they change with environmental (chemostat) conditions, can give us some initial insight on the ecological behavior of the two lytic models. 

\begin{figure}[h!]
\begin{center}
\includegraphics[width=7.5cm]{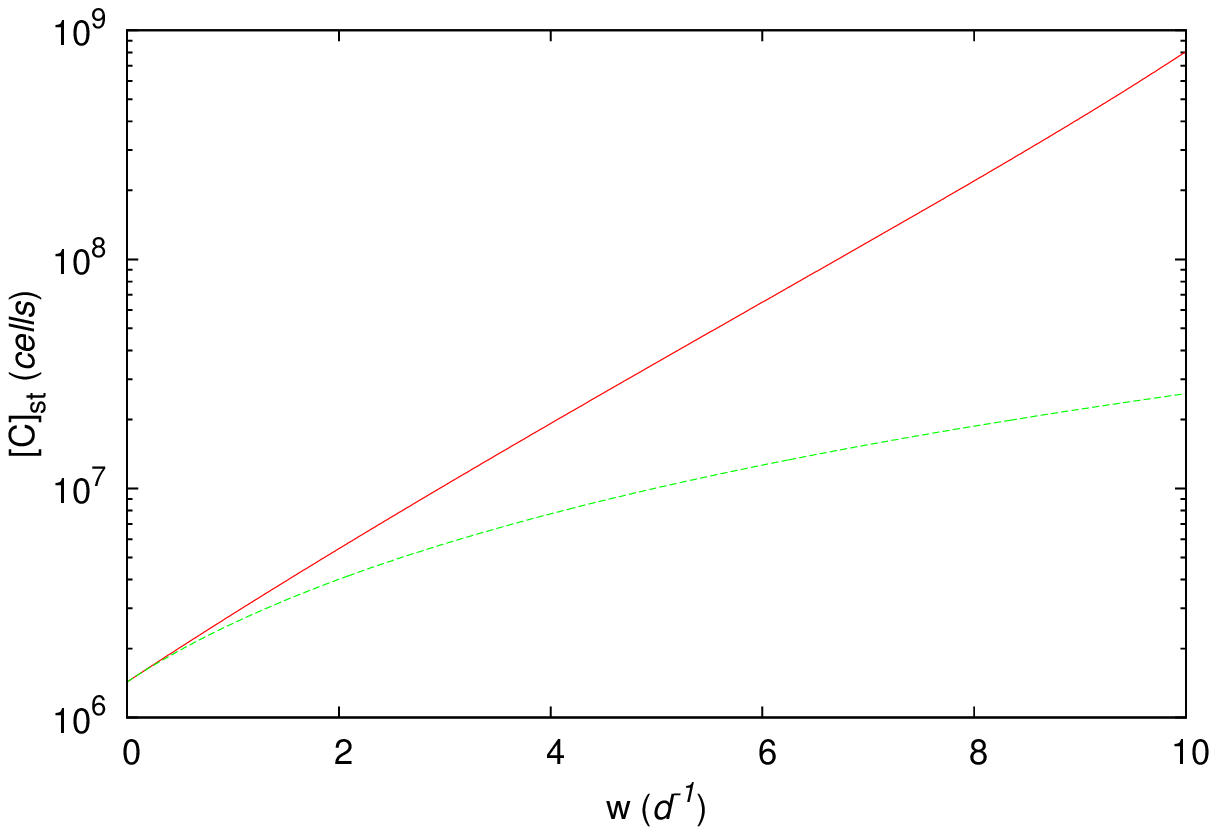}
\includegraphics[width=7.5cm]{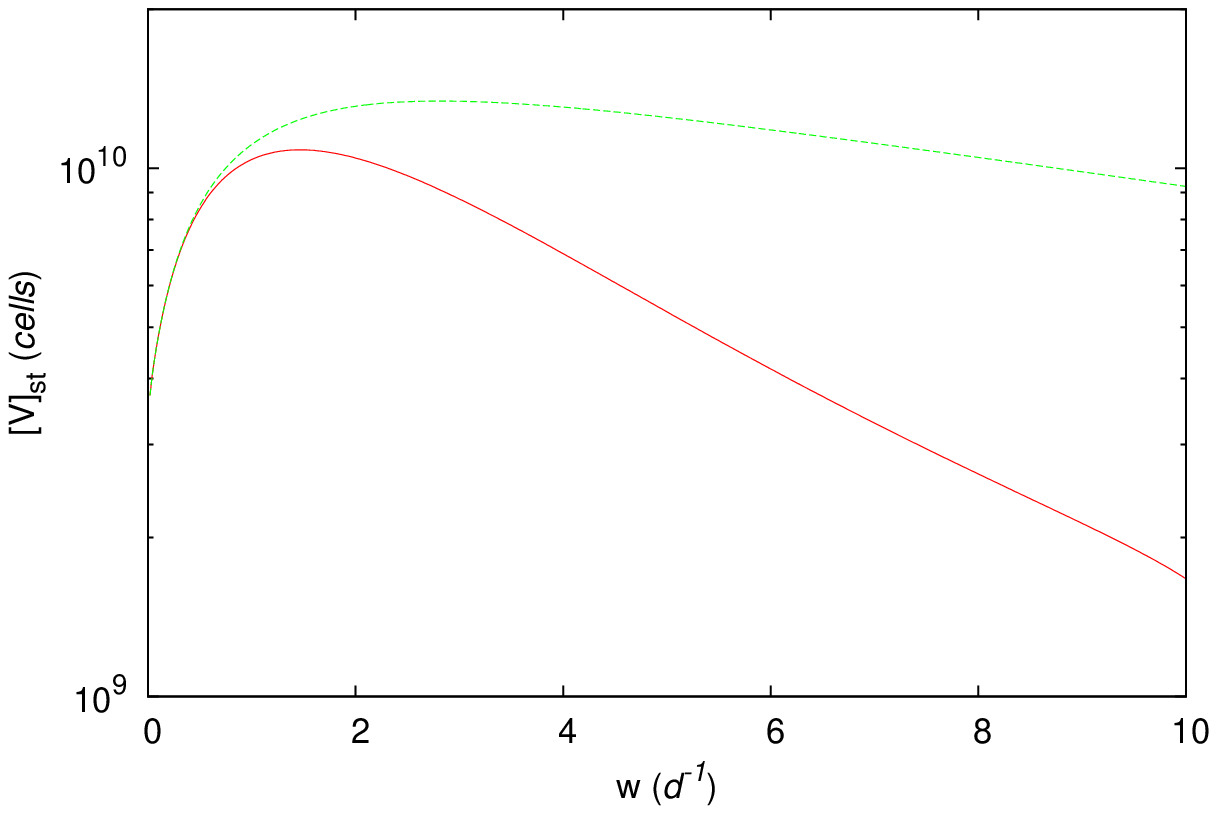}
\caption{\footnotesize{Comparison of stationary observables for the delay (red) and rate (green) lytic models at the ESS. The stationary value for the resources needed by the rate-based population is smaller than that of the delay-based one (left), whereas the population size is larger (right). For these plots, the linear shape of the trade-off $b=f(L)$ has been used, but results with the exponential form are qualitatively indistinguishable.}}
\label{all_vs_w}
\end{center} 
\end{figure}

\vspace{0.25cm}
As stated before, the ecological outcome of the two modes is qualitatively similar. For instance, for both release models, $[C]_{st}$ and $\mu_{v}$ are positively correlated with $w$ (Fig.\ref{all_vs_w}, left). $[V]_{st}$, on the other hand, shows non-monotonicity (Fig.\ref{all_vs_w}, right). These results remain valid for any of the two trade-off functions above. This positive correlation of the amount of resources needed (host cells) and viral growth rate with the dilution rate is not trivial, attending to Eqs.(\ref{stat_delayed_C}) and (\ref{DM_fitness}), and Eqs.(\ref{stat_rate_C}) and (\ref{RM_fitness}). For the host population, increasing the dilution rate increases the nutrient input rate, fostering host growth; in consequence, the viral population can grow faster as well.

\begin{figure}[h!]
\begin{center}
\includegraphics[width=7.5cm]{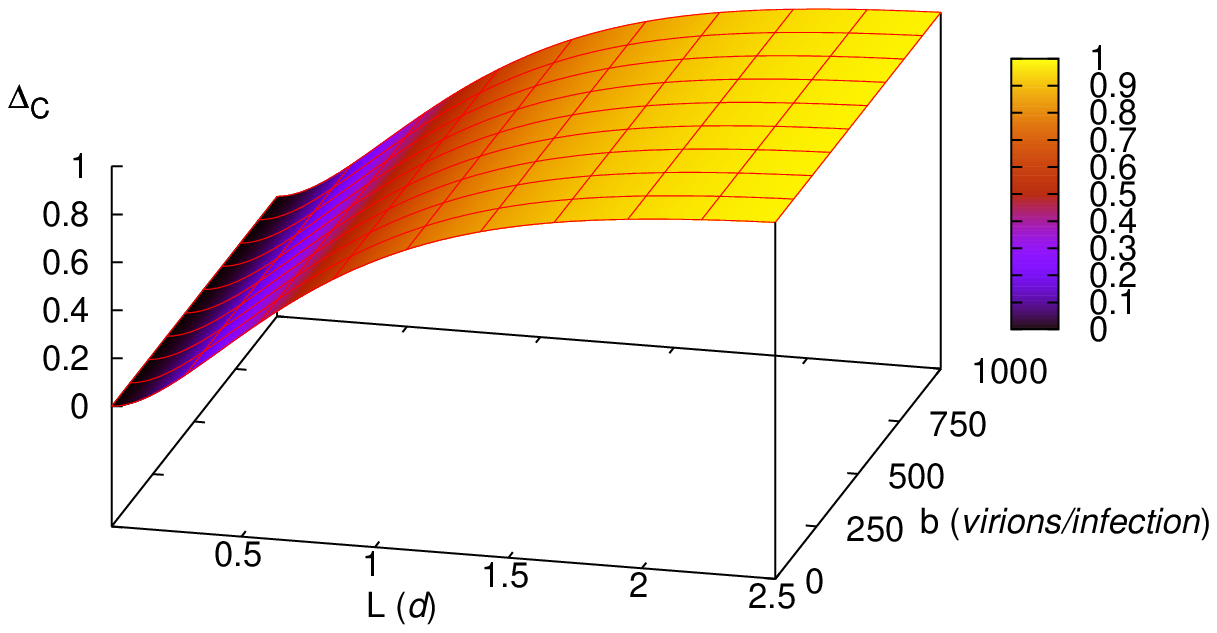}
\includegraphics[width=7.5cm]{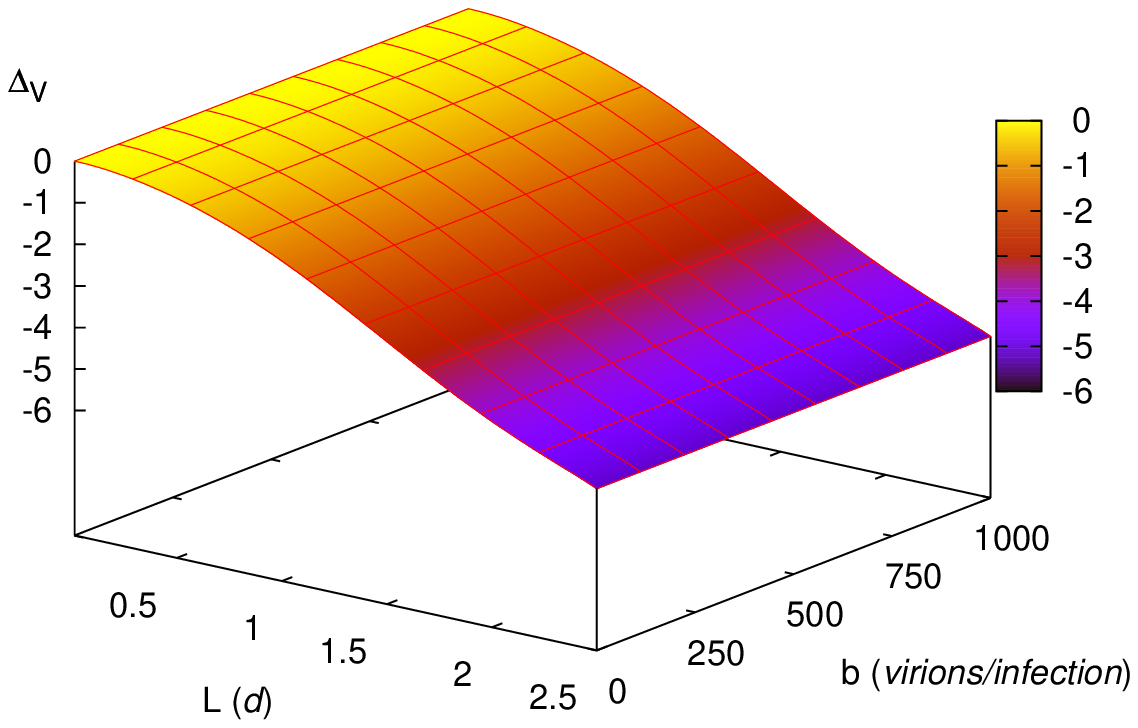}
\caption{\footnotesize{Relative difference for resources $\Delta_{C}$ (left) and population size $\Delta_{V}$ (right) between the two lytic strategies as a function of $L$ and $b=f(L)$. In these plots, the linear trade-off has been used, but the exponential one offers similar results.}}
\label{ecol_comparison}
\end{center} 
\end{figure}

\vspace{0.25cm}
\noindent
On the other hand, we can measure the relative difference $\Delta$ in the stationary concentrations of host, virus and infected cells between the two models (defined as $\Delta_{X}=1-\left[X\right]_{st_{RM}}/\left[X\right]_{st_{DM}}$, with $X=C$, $V$, or $I$, respectively), as a function of $b$ and $L$. In Fig.\ref{ecol_comparison}, we can observe that the stationary $[C]$ is lower in the case of the RM for any feasible pair of $L$ and $b$, while the stationary $[V]$ is larger. In this case, the exponential trade-off has been used, but a qualitatively similar pattern is observed for the linear form of the trade-off (results not shown).

\vspace{0.25cm}
\noindent
This improved ecological performance for the RM is a consequence of the different timing between the two lytic cycles \cite{fitness_gradient}: while in the DM there is a fixed delay between the infection of any cell and its subsequent effects (i.e. viral reproduction and disease propagation), in the RM distributed bursts have immediate consequences on the population. Accordingly, the final viral population for the RM is larger and grows faster; it also needs less resources (Fig.\ref{ecol_comparison}, and Fig.\ref{DM_V_and_muv}). According to classic competition theory \cite{tilman_book}, this would indicate that viral populations releasing continuously their offspring (e.g. shedding viruses) are better competitors for any feasible combination of $b$ and $L$. However, the fact that this strategy performs better in isolation does not necessarily mean that it can out-compete the delayed strategy when present in the same environment \cite{gene_t}. Moreover, the fact that delayed lysis and not shedding dominates in, e.g. marine environments highlights the limitation of these simplified models to produce reliable ecological and evolutionary predictions without the proper modifications \cite{fitness_gradient,new_ref_antibiotics}.

\vspace{0.25cm}
We can also define the rate of infection (ROI) as $k[V]$ \cite{eco_evo_virus}, whose behavior parallels that of the viral population size as we have assumed $k$ to be constant. For the parametrization in table \ref{table}, the frequency of infection in the population is never beyond $50-60$ infections per day. This number is reduced to $20$ infections per day when $[C_{0}]=0$ (one-stage chemostat). Such low values confirm the suitability of the single-infection assumption used here. Moreover, the difference in ROI, as $\Delta_{V}$, is negative for any realistic value of $w$, $L$ or $b$, due to the (obvious) higher speed of infection spread by the rate-based model.

\begin{figure}[h!]
\begin{center}
\includegraphics[width=7.5cm]{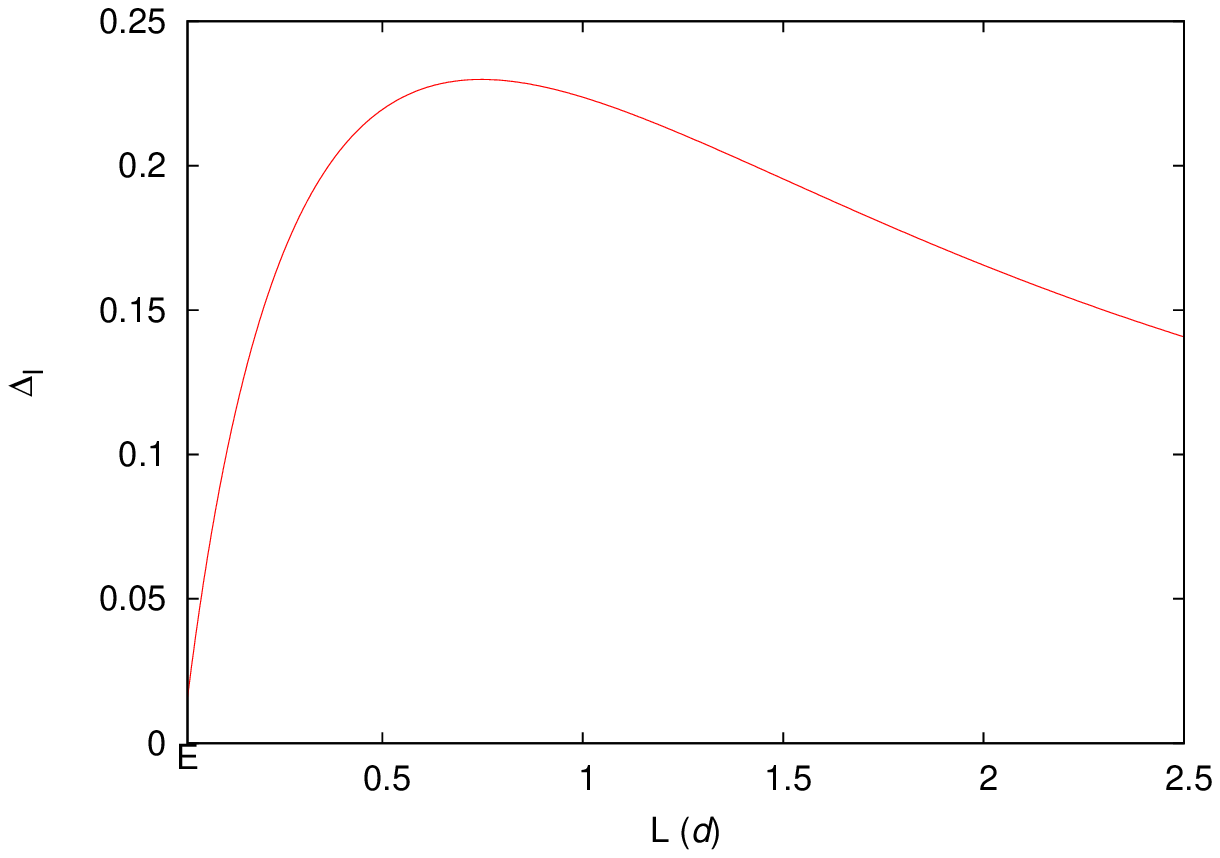}
\includegraphics[width=7.5cm]{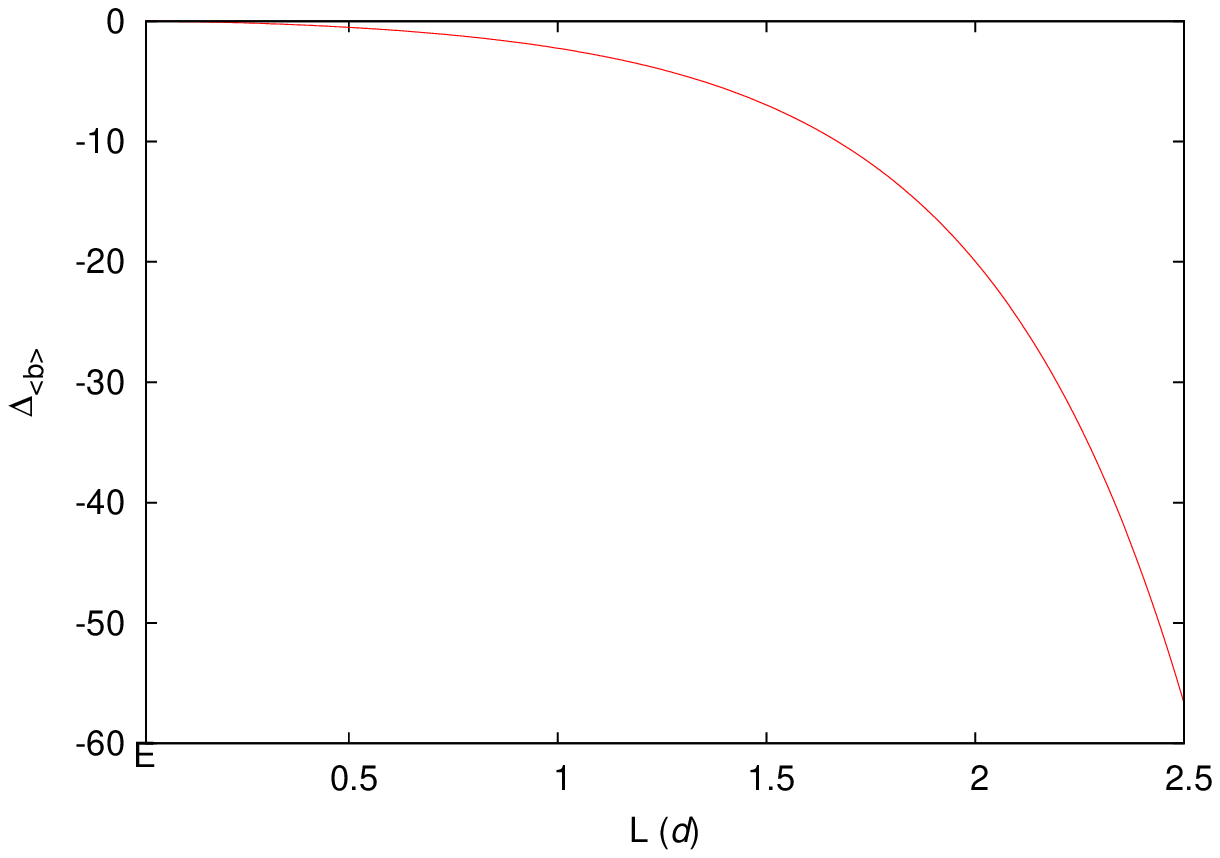}
\caption{\footnotesize{Comparison between models using as a reference the average number of surviving offspring. Left: Relative difference between the number of infected cells using the function $\Delta_{I}$ as defined in the text, with $[I]_{st}$ expressed as a function of $<b>$ (see Eq.(\ref{stat_delayed_I}) and (\ref{av_b_delay}),  and Eq.(\ref{stat_rate_I}) and (\ref{av_b_rate})). Right: Relative comparison between the surviving offspring per cell using $\Delta$ as a function of each model's $<b>$.}}
\label{newfig}
\end{center} 
\end{figure}

\vspace{0.25cm}
Lastly, we can compare the nontrivial stationary states of the two models once dilution and other mortality events are discounted from the viral offspring. It is easy to see that, when the steady-state solutions, Eq.(\ref{stat_delayed_C})-(\ref{DM_fitness}) and Eq.(\ref{stat_rate_C})-(\ref{RM_fitness}), are expressed as functions of each model's $<b>$, Eq.(\ref{av_b_delay}) and Eq.(\ref{av_b_rate}) respectively, both models offer identical values for $[C]_{st}$, $[V]_{st}$, and $\mu_{v}$, while $[I]_{st}$ is larger for the DM (Fig.\ref{newfig}, left). In other words, DM viruses need more infections to maintain growth, population density and resource requirements similar to those of the RM. However, for the same values of $(b,L)$, the RM shows a much larger amount of surviving offspring than the DM (Fig.\ref{newfig}, right), explaining why the former out-performs the latter in Figs.\ref{all_vs_w}-\ref{DM_V_and_muv}. Note that these relative differences depend only on the latent period, and therefore are not influenced by the particular trade-off assumed.

\subsection{{\it Evolutionary comparison}}
\begin{figure}[h!]
\begin{center}
\includegraphics[width=7.5cm]{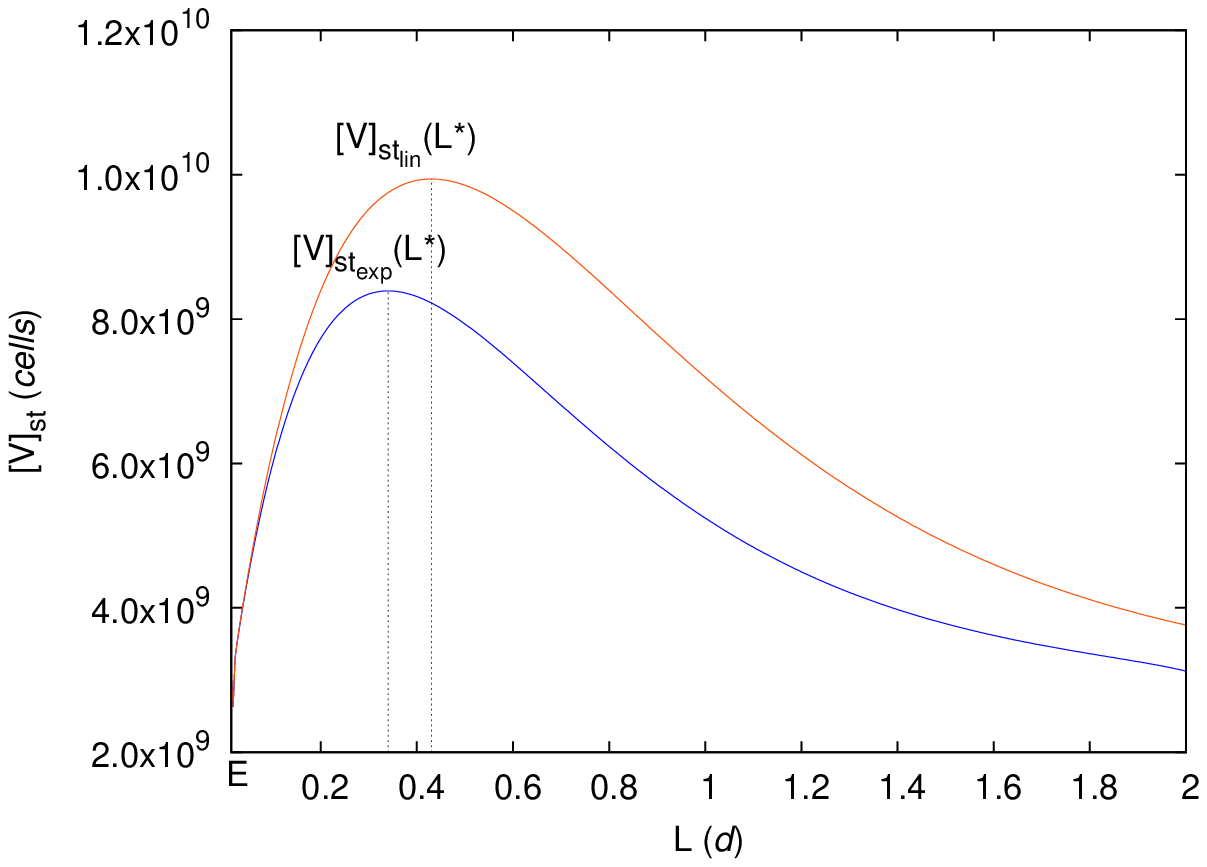}
\includegraphics[width=7.5cm]{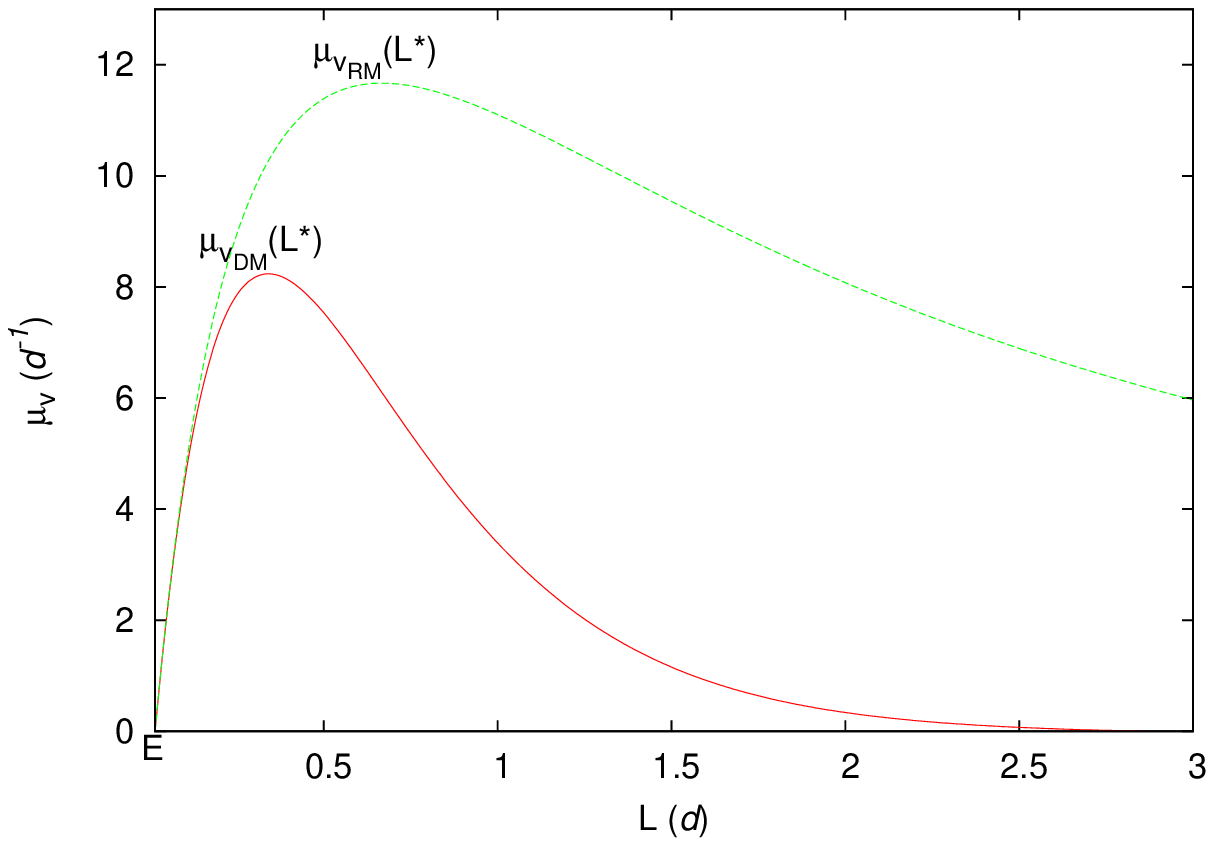}
\caption{\footnotesize{Left: Viral population size obtained with the DM using the exponential (blue) and linear (orange) forms for the trade-off $b=f(L)$. Although both observables reach a maximum at $L^{*}$, the linear trade-off yields a larger population size. Right: Fitness functions for the delay (red) and rate (green) lytic descriptions as a function of the latent period. The DM virus shows a smaller fitness, but a much steeper graph around the ESS than the RM virus. For a meaningful comparison of the fitness, a constant host population size of $10^{7}$ has been assumed.}}
\label{DM_V_and_muv}
\end{center} 
\end{figure}

Focusing now on the ESS, for both delay and rate models not only does the ESS minimize the amount of resources needed by the virus (i.e. $[C]_{st}$, Eqs.(\ref{DM_der_sol_text}) and (\ref{RM_der_sol_text})), but it also maximizes the viral population size and its fitness (Fig.\ref{DM_V_and_muv}). These results have been obtained theoretically in the past for the DM under low ROI and optimality conditions (e.g. \cite{eco_evo_virus}). This explains why the ESS coincides with the dominant phenotype in EiE experiments: the ESS makes the best use of the available resources and, thus, out-competes any other phenotype. In addition, the fitness value at the ESS increases with $[C]_{st}$ for both cycles (Eqs.(\ref{DM_fitness}) and (\ref{RM_fitness})). The hump shape shown in Fig.\ref{DM_V_and_muv} (right) has been experimentally observed \cite{linear_tradeoff}, pointing to the possibility of singular strategies in controlled environments. 

\vspace{0.25cm}
\noindent
On the other hand, the DM results in a smaller fitness than the rate-based release for any value of the latent period, including the ESS (Fig.\ref{DM_V_and_muv}, right). Importantly, the selection gradient (slope of the fitness function) close to the ESS is much larger for the delay model than for the rate model. This enhanced selection strength accelerates evolution, because it enhances differences between phenotypes \cite{AD_book}.

\vspace{0.25cm}
It is noteworthy to mention that the maturation period at the ESS, $L^{*}-E$ (step {\it iii)} of the cycle description above) is phage-independent for the DM. In this model, the time needed to assembly the new virions depends exclusively on the dilution rate and/or the host physiological state (a function of $\gamma$). In the RM, however, maturation also depends on $E$. This is due to the fact that for this model the end of the eclipse period immediately leads to a possible start of the infectious stage of the population (release of offspring). This dependence on $E$ provides the phage with control over its entire reproductive cycle.

\begin{figure}[h!]
\begin{center}
\includegraphics[width=7.5cm]{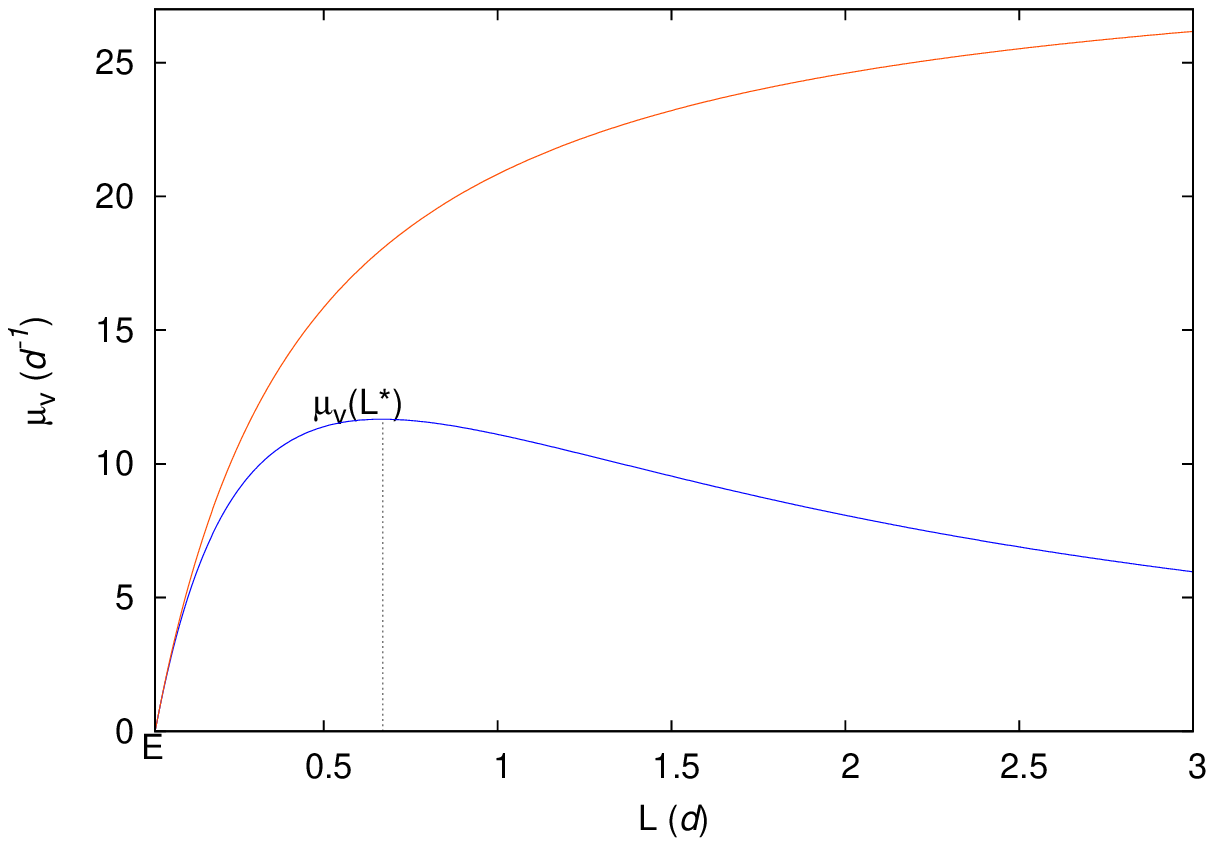}
\includegraphics[width=7.5cm]{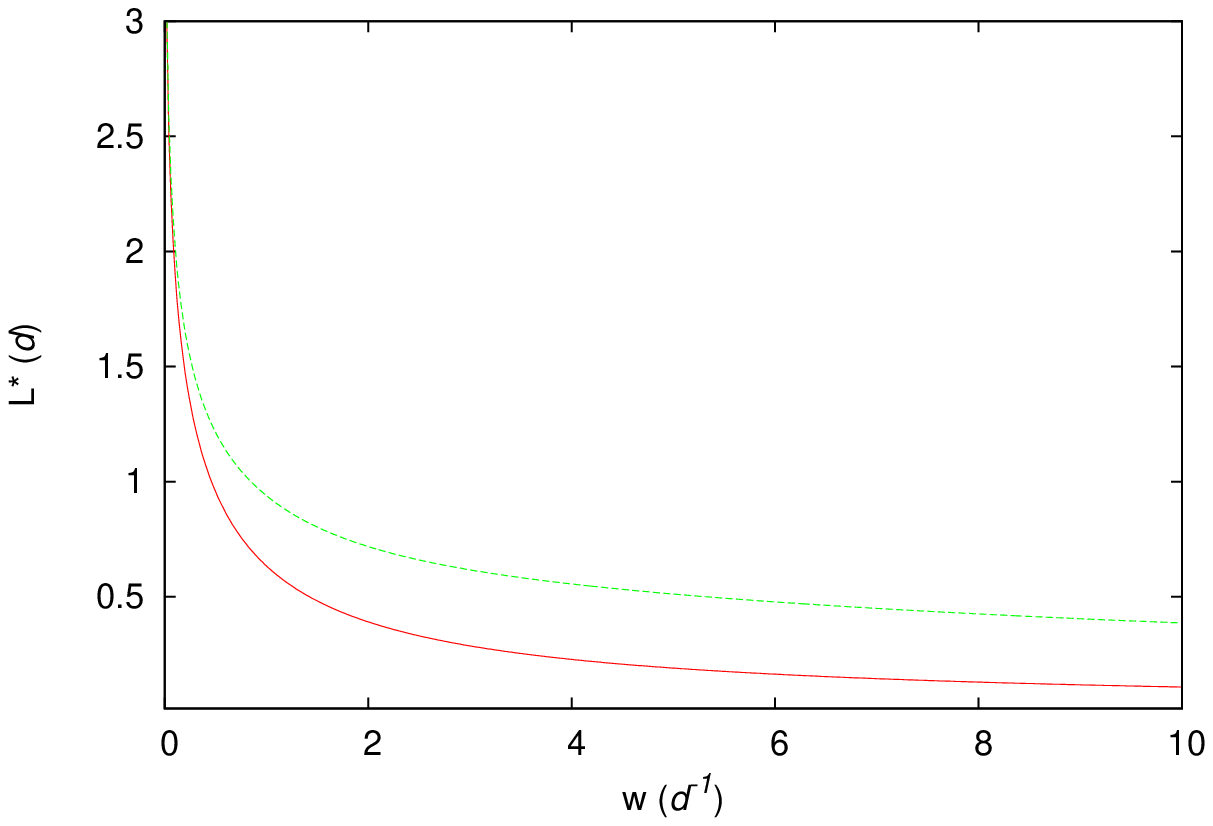}
\caption{\footnotesize{Left: Comparison of the fitness function obtained with the RM using the exponential (blue) and linear (orange) forms of the trade-off. While the exponential trade-off yields a maximum for fitness at $L^{*}$,  the fitness associated with the linear trade-off unceasingly grows with $L$. Right: Dependence of the ESS on $w$, as a proxy for host quantity, for the DM (red) and RM (green); for both strategies and trade-off forms, an improved host quantity selects for shorter latent periods. The ESS for the RM is larger for any realistic value of $w$.}}
\label{L_quantity_quality}
\end{center} 
\end{figure}

\vspace{0.25cm}
Let us delve now into the reasons for the lack of ESS in the case of the RM and linear trade-off. Mathematically, the condition for the mutant to invade, $[C_{M}]_{st}<[C_{R}]_{st}$, is translated into $b_{R}<b_{M}$ using Eqs.(\ref{stat_rate_C}) and (\ref{lin_tradeoff}); the phenotype with the larger burst size always invades. Furthermore, there is no limit to this alternation, as there is no extrema to the viral fitness (Fig.\ref{L_quantity_quality}, left). This is owing to the fact that, for a linear relationship between burst size and latent period, the fitness cost of increasing $b$ and $L$ in the RM is always smaller than the benefit. In consequence, there is no change in sign for the invasion fitness other than that expected when the roles of mutant and resident are exchanged (see Fig.\ref{PIPs}, right).

\vspace{0.25cm}
\noindent
From an evolutionary point of view, the key is again the timing of the offspring release. In the rate-based model, the trade-off between latent period and burst size influences the average number of virions liberated per unit time in the population, but offspring start to be released instantaneously. Thus, unless the resources in the host set a limit to $b$ (case of the exponential trade-off, Eq.(\ref{exp_tradeoff}) and Fig.\ref{L_quantity_quality}, left), the impact on viral fitness of increasing $b$ at the expense of increasing $L$ is always positive. Note that the fitness associated with the linear trade-off is always larger than that of the exponential one (Fig.\ref{L_quantity_quality}, left). On the other hand, increasing $(L,b)$ in the case of the delay description has a much stronger effect on viral fitness, due to the increase in the time needed to release {\it any} of the new virions \cite{fitness_gradient}. This enhanced differentiation between approaches is eventually translated into the larger selection strength observed in Fig.\ref{DM_V_and_muv} (right).

\vspace{0.25cm}
Finally, Fig.\ref{L_quantity_quality} (right) shows the behavior of $L^{*}$ when $w$, positively correlated with host quantity (see Fig.\ref{all_vs_w}), varies. For both models, a larger host availability or quality select for shorter latent periods (Fig.\ref{L_quantity_quality}, right). Thus, improved growth conditions favor shorter generation times. This result has been observed experimentally \cite{gene_t,L_w_ref} and obtained theoretically by other means (see e.g. \cite{models_solution,exponential_tradeoff}). For the RM, however, $L^{*}$ is always larger than that of the DM. Nonetheless, the continuous release allows the fitness in RM populations to be larger than that of DM ones to the smaller impact of varying the latent period for the former (Fig.\ref{DM_V_and_muv}, right).

\subsection{{\it The role of oscillations}}
As commented above, host-virus models are prone to oscillatory equilibria. Indeed, the models presented here can show such behavior by, e.g. increasing $k$ beyond $10^{-9}\,l\,cell^{-1}d^{-1}$. Oscillations prevent stationary solutions from being realized, for example due to sudden population collapse or strong fluctuations around stationary values. Thus, parametrizations leading to oscillations prevent the ecological and evolutionary analysis presented here from being applicable.

\vspace{0.25cm}
\noindent
The presence of a second chemostat (i.e. $[C_{0}]\ne0$) has helped us find a realistic region of the parameter space where oscillations are not present. Developing  an analytical framework able to predict ecological and evolutionary behavior in the presence of such oscillations remains elusive to this date.

\section{Conclusions}
The numerical value for the ESS and the qualitative behavior described here have been observed in phage experiments (tables \ref{table0} and \ref{table}). However, this is not to say that real bacteriophages are close to their (optimal) evolutionarily stationary state. The intra- and inter-specific variability observed in measured latent periods indicates that there are other factors, not considered in this idealized study, that contribute to the exact value of the latent period shown by real viruses. For instance, constantly changing environmental conditions that include hosts potentially co-evolving with the virus will most certainly change the latent period\footnote{Among other traits, for instance $k$.} selected for in each case. 

\vspace{0.25cm}
\noindent
This is especially true for marine environments, for which strict stationary conditions are hardly found even in stratified waters. Keeping the same formalism used here, we can implement easily more realistic environments changing Eq.(\ref{nutrient_equation}) (for instance, incorporating remineralization of nutrients by bacteria, or periodic pulses), and playing with the sources of ``fresh'' host or dilution. Nonetheless, the study under chemostat conditions presented here provides valuable information on the qualitative behavior of different lytic strategies.

\vspace{0.25cm}
Continuous cultures are very extended in the experimental literature. As shown here, in these environments populations with distributed latent periods prove to have many ecological and evolutionary advantages. Everything being equal, RM populations start releasing offspring (i.e. become infectious) earlier, providing the RM virus with competitive advantage over the DM one. Moreover, dynamic evolution of the latent period  under the same environmental conditions leads the rate-based population toward an ESS in which the virus utilizes less resources to produce larger population sizes and larger viral fitness with smaller generation times.

\vspace{0.25cm}
\noindent
The strength of selection for this model is, however, much smaller than for the case of DM viruses. Smaller fitness gradients slow down evolutionary succession and, thus, may prevent the ESS for RM viruses from being eventually realized. Moreover, it can in principle put, e.g. shedding strategies in competitive disadvantage if the bacterial host co-evolves, because a burst lytic virus may adapt more quickly to changes in the host. Even in the absence of DM viruses, RM viruses may not adapt quickly enough to host co-evolution, leading to a much slower (or even eventually vanishing) Red-Queen dynamics. However, the expected changing fitness landscape will influence non-trivially the adaptation rate in both descriptions. For viruses that can show either infection cycle, shedding provides quick invasion whereas burst lysis provides quick adaptation.

\vspace{0.25cm}
On the other hand, we have mathematically proved that the linear form for the trade-off between burst size and latent period yields an endless evolutionary succession in the case of the RM. Linear trade-offs have been experimentally observed for bacteriophages, and maxima for the viral fitness have been measured in the same experiments. Therefore, the standard formulation of the RM presented here may lead to unrealistic predictions about bacteriophage evolution in steady environments. Thus, if used to describe long-term behavior of this and other phages such as phages with non-linear virion assemblage or, e.g. shedding, more appropriate environments, biological constraints, and trade-offs need to be used. Possible improvements to capture a shedding cycle correctly should include an explicit eclipse period in the exponential release distribution, and de-couple offspring release from cell mortality. On the other hand, further research is needed in order to find which functional forms for the relation between life-history traits correspond specifically to marine phages.

\vspace{0.25cm}
Thus, the classic modeling trade-off between simplicity and realism materializes once again. The study presented here shows that, with the trade-offs analyzed, even though applying the simpler RM model to bacteriophages provides qualitatively similar ecological results to those of the DM, the former may not be reliable for evolutionary matters, leaving the latter as the only available alternative. On the other hand, the delay terms complicate the analysis of the DM. This prevents its inclusion in bigger modeling frameworks such as models for oceanic biogeochemistry, which lack a realistic explicit representation of marine viruses. Thus, a new theoretical description is needed, able to capture the essential ecological {\it and} evolutionary aspects of the marine host-virus dynamics without the use of delay terms. Finding new paths to the regularization of the delay terms fulfilling these features remains as an open question.

\vspace{0.25cm}
Another open issue relates to finding a theoretical framework able to tackle the eco-evolutionary interactions in host-virus systems. In this paper, we focused on the analytical description of stationary states in both ecological and evolutionary timescales, ensured by the chosen environmental conditions and parametrization. However, such a framework could describe analytically more realistic situations such as the evolutionary succession (i.e. transients) observed in our simulations. This theoretical framework would be able to capture the feedback loop between ecology and evolution provided by rapid evolutionary events, non-vanishing evolutionary jumps, and overlapping generations. These three features break in one way or another the simplifying assumptions of the available theoretical frameworks such as adaptive dynamics.

\vspace{0.25cm}
This study evidences the importance of taking into account {\it both} ecological and evolutionary aspects of the dynamics between host and virus, subject to rapid evolution. This is especially relevant if we are interested in reliable long-term predictions for the system under scrutiny. The inclusion of viruses in the description of biogeochemical cycles is one important example, as reliable estimates of viral dynamics are crucial to understand any future climate change scenario. Another sound example is phage therapy, which is re-emerging as an alternative to antibiotics. The design of efficient treatments requires a reliable estimate not only of instantaneous population sizes but also of possible  co-evolutionary events between phages and bacteria. The new theoretical alternatives suggested here will prove to be essential to this end.

\section*{Acknowledgments}
We want to thank Anne Maria Eikeset, Yoh Iwasa, Roger Kouyos, Duncan Menge, Alex Washburne, and Joshua Weitz for helpful discussions in early stages of this research. We thank the editor and the two anonymous reviewers for their feedback, which has improved enormously this manuscript. We are also grateful for the support from NSF under grant OCE-1046001 and by the Cooperative Institute for Climate Science (CICS) of Princeton University and the National Oceanographic and Atmospheric Administration's (NOAA) Geophysical Fluid Dynamics Laboratory (GFDL).

\setcounter{section}{0}
\renewcommand{\thesection}{\Alph{section}}

\clearpage
\section{Appendix A: Table of definitions and parameter values}
\label{AppA}

\setcounter{table}{0}
\renewcommand{\thetable}{\Alph{table}}

\begin{table*}[h!]
\caption{\footnotesize{Compilation of symbols and parameter values. Data for the host taken from \cite{host_data}. Data for the virus into the ranges used/shown in \cite{gene_t,eco_evo_virus,joshua_PNAS,exponential_tradeoff,joshua_other,realistic_values}. The yield parameter, $Y$, has been adjusted to obtain a maximum growth $\mu_{max}=Y\;V_{max}=18 d^{-1}$, value which is also in harmony with the previous references.}}
\vskip4mm
\centering
{\footnotesize
\begin{tabular}{|c|c|c|c|}
\hline
Symbol&Description&Units&Value\\
\hline\hline
$[N]$&Dissolved Inorganic Nitrogen Concentration&$mol\;l^{-1}$&Variable\\
\hline
$[C]$&Non-infected Host Concentration&$cell\;l^{-1}$&Variable\\
\hline
$[I]$&Infected Host Concentration&$cell\;l^{-1}$&Variable\\
\hline
$[V]$&Free Virus Concentration&$cell\;l^{-1}$&Variable\\
\hline
$k_{L}$&Lysis Rate&$d^{-1}$&Evolutionary variable\\
\hline
$L$&Latent Period&$d$&Evolutionary variable\\
\hline
$b$&Burst Size&$virions$&Evolutionary variable\\
\hline
$\mu_{v}$&Virus Population Growth Rate&$d^{-1}$&Variable\\
\hline
$\mu$&Host Population Growth Rate&$d^{-1}$&Variable\\
\hline
\hline
$\mu_{max}$&Maximum Host Population Growth Rate&$d^{-1}$&$18$\\
\hline
$Y$&Yield Parameter&$cell\;mol^{-1}$&$4.5\times10^{13}$\\
\hline
$V_{\max_{N}}$&Maximum Nutrient Uptake Rate&$mol\;cell^{-1}\;d^{-1}$&$4\times10^{-13}$\\
\hline
$K_{N}$&Half-Saturation Constant the Nutrient&$mol\;l^{-1}$&$10^{-6}$\\
\hline
$k$&Adsorption Rate&$l\;cell^{-1}\;d^{-1}$&$10^{-10}$, $5\times10^{-9}$\\
\hline
$m$&Virus Mortality Rate&$d^{-1}$&$0.1$, $5$\\
\hline
$M$&Maturation Rate&$virions\;d^{-1}$&$1.44\times10^{3}$\\
\hline
$E$&Eclipse Period&$d$&$0.0139$\\
\hline
$\gamma$&Host Resources Decay Rate&$d^{-1}$&$1.44$\\
\hline
$[C_{0}]$&Non-infected Host Supply Concentration&$cell\;l^{-1}$&$10^{8}$\\
\hline
$[N_{0}]$&Dissolved Inorganic Nutrient Supply Concentration&$mol\;l^{-1}$&$50\times10^{-6}$\\
\hline
$w$&Chemostat Dilution Rate&$d^{-1}$&$2.4$\\
\hline
\end{tabular}
}
\label{table}
\end{table*}

\section{Appendix B}
\label{AppB}

\subsection{Feasibility of the trivial stationary states}
For both models, a trivial stationary solution is given by the phage-free state $[V]_{st}=[I]_{st}=0$, $[C]_{st}=w[C_{0}]/(w-\overline{\mu})$, with $\overline{\mu}=\mu([N]_{st})$. Stationary conditions also require that:

\begin{eqnarray}
\dfrac{d[N]}{dt}=0&\Longleftrightarrow\left([N_{0}]-[N]_{st}\right)=&\overline{\mu}\dfrac{[C]_{st}}{Y}\\
&&=\overline{\mu}\dfrac{w[C_{0}]}{Y\left(w-\overline{\mu}\right)},
\label{Nstat1}
\end{eqnarray}

\noindent
or, rearranging terms:

\begin{equation}
\left([N_{0}]-[N]_{st}\right)\left(w-\overline{\mu}\right)=\overline{\mu}\dfrac{[C_{0}]}{Y}
\label{Nstat2} 
\end{equation}

\noindent
Because $\overline{\mu}=\mu([N]_{st})$ increases with $[N]_{st}$ and vanishes at zero, and the two terms on the left hand side decrease with $[N]_{st}$ and are positive at zero, the Intermediate Value Theorem ensures a solution to the equation above for which both sides are positive (i.e. $0<\mu([N]_{st})<w$ and $0<[N]_{st}<[N_{0}]$. Thus, the trivial stationary state is always feasible.

\subsection{Feasibility of the non-trivial solutions}
For both models, an analysis similar to the one above ensures that $0<\mu([N]_{st})$ and $0<[N]_{st}<[N_{0}]$. On the other hand, the necessary positivity condition for $[C]_{st}$, $[V]_{st}$, and $[I]_{st}$ imposes:

\begin{equation}
\mu([N]_{st})\ge w\left(1-\dfrac{[C_{0}]}{[C]_{st}}\right).
\end{equation}

\noindent
and, $L\le\ln{b}/w$ for the DM, whereas for the RM the latter condition becomes $L\le(b-1)/w$.

\subsection{Invasion analysis}

\subsubsection{{\it Delay Model:}}
 If we assume that the invader perturbs the otherwise stable state of the resident population (subindex $R$), the dynamic equations for the mutant (subindex $M$) can be written as:

\begin{eqnarray}
\dfrac{d[V_{M}](t)}{dt}&=&\left(b_{M}\;k[C_{R}]_{st}[V_{M}]_{t-L_{M}}\right)e^{-wL_{M}}-k[C_{R}]_{st}[V_{M}]-m[V_{M}]-w[V_{M}],\label{delayed_VM_eq}\\
\dfrac{d[I_{M}](t)}{dt}&=&k[C_{R}]_{st}[V_{M}]-\left(k[C_{R}]_{st}[V_{M}]_{t-L_{M}}\right)e^{-wL_{M}}-w[I_{M}]\label{delayed_IM_eq},
\end{eqnarray}

\noindent
where $[C_{R}]_{st}$ ($[C_{M}]_{st}$) represents Eq.(\ref{stat_delayed_C}) calculated using the resident (mutant) traits. By definition, the evolutionarily stable strategy (ESS) cannot be invaded by any mutant or immigrant phenotype. Thus, the sign of the invasion eigenvalue (i.e. eigenvalue associated with the equations above) will provide the conditions for the phenotype ($L^{*}$, $b^{*}$) to be uninvadable. The eigenvalues $\lambda$ are the result of solving $|J + J_{D}e^{-\lambda L_{M}}-\lambda I|=0$ \cite{DM_stability}, where $J$ is the Jacobian matrix associated with the instantaneous terms of the equations, $J_{D}$ that of the delayed terms, and $I$ is the identity matrix. The condition above can be translated into:

\begin{equation}      
\begin{vmatrix}
k[C_{R}]_{st}\left(b_{M}\;e^{-(w+\lambda)L_{M}}-1\right)-w-m-\lambda&0\\
-k[C_{R}]_{st}\left(e^{-(w+\lambda)L_{M}}-1\right)&-w-\lambda
\end{vmatrix}=\left(A(\lambda)-\lambda\right)\left(B-\lambda\right)=0.
\label{delayed_J_condition2}
\end{equation}

\noindent
One eigenvalue is, trivially, given by $\lambda=B=-w$. Thus, if the other eigenvalue, resulting from solving the implicit equation $\lambda=A(\lambda)=k[C_{R}]_{st}\left(b_{M}\;e^{-(w+\lambda)L_{M}}-1\right)-w-m$ is positive, the mutant can invade, whereas a negative value will ensure unbeatability for the resident. This remaining eigenvalue is given by:

\begin{equation}
\lambda=\dfrac{1}{L_{M}}W_{n}\left(k[C_{R}]_{st}b_{M}L_{M}e^{-(k[C_{R}]_{st}+m)L_{M}}\right)-\left(k[C_{R}]_{st}+w+m\right),
\label{DM_eigen_sol} 
\end{equation}

\noindent
where $W_{n}(z)$ is the so-called Lambert function, defined as the solution to $W_{n}(z)e^{W_{n}(z)}=z$ \cite{lambert}. The condition $\lambda=0$ provides the marginal case:

\begin{equation}
A\left(0\right)=0\;\;\;\;\;\;\Longleftrightarrow\;\;\;\;\;\;\dfrac{m+w}{k(b_{M}e^{-wL_{M}}-1)}=[C_{R}]_{st}\;\;\;\;\;\;\Longleftrightarrow\;\;\;\;\;\;[C_{M}]_{st}=[C_{R}]_{st},
\label{DM_C_condition} 
\end{equation}

\noindent
For the resident to resist invasion (i.e. $\lambda<0$), $[C_{R}]_{st}<[C_{M}]_{st}$. Alternatively, for the mutant to invade ($\lambda>0$), $[C_{M}]_{st}<[C_{R}]_{st}$. Therefore, the phenotype that minimizes $[C]_{st}$ will be an ESS candidate. Thus:

\begin{eqnarray}
\dfrac{d[C]_{st}}{dL} = 0 &\Longleftrightarrow&\dfrac{db}{dL}=w\;b;\nonumber\\\nonumber\\&\Longleftrightarrow& f'\left(L^{*}\right)=w\;f\left(L^{*}\right).
\label{DM_der_sol}
\end{eqnarray}

\noindent
This condition, deduced in \cite{models_solution} by other means, can also be deduced looking for the invading strategy that maximizes $\mu_{v}$, Eq.(\ref{DM_fitness}). From Eq.(\ref{DM_der_sol}), it easily follows that $d^{2}[C]_{st}/dL^{2}$ is positive at $L^{*}$. Thus, the solution to Eq.(\ref{DM_der_sol}) indeed provides an uninvadable strategy that maximizes fitness, i.e. the ESS.

\vspace{0.25cm}
Equivalently, the condition given by Eq.(\ref{DM_der_sol}) can be found by realizing that Eq.(\ref{DM_eigen_sol}) can be used to define the invasion fitness function, $s_{L_{R}}(L_{M})=\lambda$. Thus, evolutionary singularities are given by the points $L^{*}$ such that the derivative of $s$ with respect to $L$ vanishes, $\frac{\partial s_{L_{R}}(L_{M})}{\partial L_{M}}|_{L^{*}}=0$. It is also possible to show that both the second derivative of $s_{L_{R}}(L_{M})$ and $\frac{d}{dL}\left(\frac{\partial s_{L_{R}}(L_{M})}{\partial L_{M}}\left.\right|_{L_{M}=L_{R}=L}\right)$ are, at least for the parametrization in table \ref{table}, negative when evaluated at $L=L^{*}$. In this way, the analytical conditions for $L^{*}$ to be an ESS and a CSS, respectively, are fulfilled.

\subsubsection{{\it Rate Model:}}
We now follow similar steps to deduce the expressions for the ESS in the case of the rate model. The equations for the mutant are, in this case:

\begin{eqnarray}
\dfrac{d[V_{M}](t)}{dt}&=&b_{M}\;k_{L,M}[I_{M}]-k[C_{R}]_{st}[V_{M}]-m[V_{M}]-w[V_{M}]\label{rate_VM_eq}\\
\dfrac{d[I_{M}](t)}{dt}&=&k[C_{R}]_{st}[V_{M}]-k_{L,M}[I_{M}]-w[I_{M}]\label{rate_IM_eq},
\end{eqnarray}

\noindent
and the characteristic equation for the invasion eigenvalue is given by:

\begin{equation}
|J-\lambda I|=
\begin{vmatrix}
-k_{L,M}-w-\lambda&k[C_{R}]_{st}\\
k_{L,M}b_{M}&-k[C_{R}]_{st}-w-m-\lambda
\end{vmatrix}=
\begin{vmatrix}
A-\lambda&B\\
C&D-\lambda
\end{vmatrix}=0.
\label{rate_J_condition}
\end{equation}

\noindent
$A$ and $D$ are by definition negative in any feasible scenario. Thus, the only remaining condition to be fulfilled for the resident state to be uninvadable is, following the Routh-Hurwitz criteria, $BC<AD$. After some algebra, this condition is translated again into $[C_{R}]_{st}<[C_{M}]_{st}$. Therefore, the phenotype minimizing $[C]_{st}$ will be a possible ESS. Thus, using Eq.(\ref{stat_rate_C}):

\begin{eqnarray}
\dfrac{d[C]_{st}}{dL} = 0 &\Longleftrightarrow&\dfrac{db}{dL}=\dfrac{w\;b}{1+w\;L};\nonumber\\\nonumber\\&\Longleftrightarrow& f'\left(L^{*}\right)=\dfrac{w\;f\left(L^{*}\right)}{1+w\;L^{*}}.
\label{RM_der_sol}
\end{eqnarray}

\noindent
This same condition can be reached by defining the invasion fitness function $s_{L_{R}}(L_{M})=BC-AD$ and using $\frac{\partial s_{L_{R}}(L_{M})}{\partial L_{M}}|_{L^{*}}=0$.

\clearpage
\section*{Figures}

\end{document}